\documentclass[a4paper,11pt]{article}
\pdfoutput=1 

\usepackage{jheppub} 

\usepackage[T1]{fontenc} 

\def\be{\begin{equation}}
\def\ee{\end{equation}}
\def\bea{\begin{eqnarray}}
\def\eea{\end{eqnarray}}
\def\ben{\begin{equation}}
\def\een{\end{equation}}
\def\bean{\begin{eqnarray}}
\def\eean{\end{eqnarray}}

\def\pd{\partial}
\def\a{\alpha}
\def\b{\beta}
\def\g{\gamma}
\def\d{\delta}
\def\m{\mu}
\def\n{\nu}
\def\t{\tau}

\def\l{\lambda}

\def\r{\rho}

\def\bg{\bar{g}}

\def\bD{\bar{D}}

\def\bp{\bar{\phi}}
\def\s{\sigma}
\def\e{\epsilon}

\def\bma{\begin{pmatrix}}
\def\ema{\end{pmatrix}}

\def\f{\phi}

\def\bf{\bar{\phi}}
\def\Ep{\xi}

\def\D{\nabla}
\def\bD{\bar{\nabla}}
\def\bg{\bar{g}}
\def\bi{\begin{itemize}}
\def\ei{\end{itemize}}
\def\bn{\bar{\nabla}}
\def\br{\bar{R}}
\def\bp{\bar{\phi}}

\title{\boldmath Conformal and non Conformal  Dilaton Gravity.}

\author[a,1]{Enrique Alvarez,\note{Corresponding author.}}
\author[a]{Mario Herrero-Valea,}
\author[b]{C.P. Mart\'{\i}n}

\affiliation[a]{Departamento de F\'{\i}sica Te\'orica and Instituto de F\'{\i}sica Te\'orica, IFT-UAM/CSIC\\Universidad Aut\'onoma, 20849 Madrid, Spain}
\affiliation[b]{Universidad Complutense de Madrid (UCM), Departamento de F\'{\i}sica Te\'orica I \\
Facultad de Ciencias F\'{\i}sicas, Av. Complutene S/N (Ciudad Univ.)\\
E-28040 Madrid, Spain}

\emailAdd{enrique.alvarez@uam.es}
\emailAdd{mario.herrero@csic.es}
\emailAdd{carmelop@fis.ucm.es}

\abstract{
The quantum dynamics of the gravitational field non-minimally coupled to an (also dynamical) scalar field is studied in the {\em broken phase}. For a particular value of the coupling the system is classically conformal, and can actually be understood  as the group averaging of Einstein-Hilbert's action under conformal transformations.
Conformal invariance implies a simple Ward identity asserting that  the trace of the equation of motion for the graviton is the equation of motion of the  scalar field.
We perform an explicit one-loop computation to show that the DeWitt effective action is not UV divergent {\em on shell} and to find that the Weyl symmetry Ward identity is preserved {\em on shell} at that level. We also discuss the fate of this Ward identity at the two-loop level --under the assumption that the two-loop UV divergent part of the effective action can be retrieved from the Goroff-Sagnotti counterterm-- and
 show that its preservation in the renormalized theory requires the introduction of counterterms which exhibit a logarithmic dependence on the dilaton field.
}
\begin{document}
{\flushright{IFT-UAM/CSIC-14-022, ~~FTUAM-14-10, ~~FTIUCM-17-2014}}

\maketitle
\flushbottom

\newpage
\section{Introduction.}
It is a cherished belief that some sort of scale invariance should be relevant when studying physics at very short distances. In flat space-time is always possible to get an (improved) traceless energy-momentum tensor
\be
T\equiv \left.g^{\m\n}{\d S_{\text{matt}}\over \d g^{\m\n}}\right|_{g_{\m\n}=\eta_{\m\n}}=0
\ee

The cosmological constant is then related to the trace of the gravitational equations of motion (EM)
\be
\left.\Lambda\equiv g^{\m\n}{\d S \over \d g^{\m\n}}\right|_{g_{\m\n}=\eta_{\m\n}}=\left.\left(1-{n\over 2}\right)R-n\l\right|_{g_{\m\n}=\eta_{\m\n}}=-n\l=0
\ee

This means that exact conformal invariance prevents a cosmological constant \cite{Pisarski}. This is a well-known fact, which undoubtly can be traced back much earlier than we were able to do. This is a strong physical motivation to further study these theories.

 When gravitational fields are dynamical, the corresponding symmetry is Weyl invariance (we shall understand conformal symmetry always in this sense), local rescalings of the spacetime metric. Indeed {\em conformal (super)gravity} \cite{FT} is such a  theory in which  Weyl invariance is implemented in four spacetime dimensions starting with the lagrangian\footnote{
Let us agree once and for all to denote in future formulas the riemannian volume element as
\be
d(vol)\equiv \sqrt{|g|}~d^n x
\ee
}
\be
L\equiv \sqrt{-g}~W_{\m\n\r\s}~W^{\m\n\r\s}\equiv \sqrt{|g|}~W_4
\ee
where $W_{\m\n\r\s}$ is  Weyl's tensor, the tracefree piece of Riemann's tensor. It is explicitily defined in terms of the Riemann tensor as
\bea\label{weylten}
&&W_{\m\n\r\s}\equiv R_{\m\n\r\s}-{1\over n-2}\left(g_{\m\r} R_{\n\s}- g_{\m\s} R_{\n\r} - g_{\n\r} R_{\m\s} + g_{\n\s} R_{\m\r}\right)+\nonumber\\
&&+{1\over (n-1)(n-2)}R\left(g_{\m\r}g_{\n\s}-g_{\m\s}g_{\n\r}\right)
\eea

then
\be
W_4\equiv R_{\m\n\r\s}^2 -{4\over n-2}  R_{\m\n}^2+{2\over \left(n-1\right)\left(n-2\right)} R^2.
\ee

This lagrangian is point conformally invariant under
\be
\widetilde{g}_{\m\n }(x)\equiv \Omega^2(x) g_{\m\n}(x)
\ee
(this means by definition that the gravitational field has conformal weight $w=-2$) in four  dimensions only. There are local invariants in arbitrary dimensions, involving derivatives of the Weyl tensor and the Fefferman-Graham obstruction, whose existence is guaranteed, but which is not known explicitly in general \cite{FeffermanG}. It would be interesting to study the physics of actions based on the integral of those invariants, but we shall refrain from doing so in this paper.
\par
 It has been argued that conformal supergravities can be  finite at the quantum level provided they have enough supersymmetry. Nevertheless, there is always some tension, at least at the perturbative level, with unitarity, because the propagator is quartic in the momentum, which implies ghost excitations and/or tachyon behavior. It is actually not clear in spite of some insightful attempts  whether a non-perturbative unitary definition of the theory is possible at all.
\par
It is nevertheless quite easy to construct a much simpler {\em conformal dilaton gravity} (CDG) free of these problems by the procedure of {\em group averaging}, that is, perform a conformal transformation on the Einstein-Hilbert lagrangian and promote the Weyl rescaling factor to the status of a new field. It seems that the first to consider CDG was Dirac \cite{Dirac} in a very interesting paper in which he related the {\em large numbers hypothesis} with the old unified  theory of Hermann Weyl. Other interesting pioneering works on this theory include \cite{Englert}\cite{Fradkin}\cite{Jackiw}. In those works CDG was considered as a conformally invariant off-mass shell extension of quantum gravity in the context of the early attempts to understand the physical meaning\footnote{In Duff's words {\em real Weyl invariance has anomalies; pseudo-Weyl invariance (i.e. involving a spurion) does not. This is a regularization-scheme-independent statement.} It remains of course to decide which metric couples to which matter.}
of the conformal anomaly \cite{Duff}.

Under a Weyl rescaling \footnote{We use the Landau-Lifshitz spacelike conventions.
The flat tangent metric is
\be
\eta_{ab}\equiv diag\,\left(1,-1,-1,-1\right)
\ee
The Riemann tensor reads
\be
R^\m\,_{\n\a\b}\equiv \pd_\a \Gamma^\m_{\n\b}-\pd_\b \Gamma^\m_{\n\a}+
\Gamma^\m_{\sigma\a}\Gamma^{\sigma}_{\n\b}-\Gamma^\m_{\sigma\b}\Gamma^\sigma_{\n\a}
\ee
and  the Riccci tensor 
\be
R_{\m\n}\equiv R^\l\,_{\m\l\n}
\ee
The Einstein-Hilbert action is defined as
\be
S=-\frac{c^3}{2\kappa^2}\int \sqrt{|g|}\left(R-2\lambda\right)+S_{matter}
\ee
with $\kappa^2\equiv 8\pi G$.
}the Einstein-Hilbert lagrangian behaves as
\be
\sqrt{|\widetilde{g}|}\widetilde{R}=\sqrt{|g|}\bigg[\Omega^{n-2}R+(n-1)(n-2)\Omega^{n-4}(\nabla\Omega)^2\bigg]
\ee

Where we have neglected a total derivative which yields a boundary term. We then define a {\em gravitational scalar} field through
\be
\Omega\equiv {1\over M_p}\left({(n-2)\over 4(n-1)}\right)^{1\over n-2}~\phi^{2\over n-2}
\ee
(where the n-dimensional Planck mass is defined as $M_p^{n-2}\equiv{1\over 16\pi G_n}$)
obtaining  the CDG lagrangian
\be\label{Scdg}
S_{CDG}=\int d(vol)~\left(-{n-2\over 8(n-1)}~R~ \phi^2-{1\over 2}g^{\m\n}\nabla_\m\phi\nabla_\n \phi\right)
\ee

Classically, CDG   reduces to General Relativity (GR) in the gauge
\be
\phi=\sqrt{8(n-1)\over n-2} M_p^{n-2\over 2}
\ee
which is of course only accessible as long as we are in the broken phase; and to {\em unimodular gravity} \cite{Alvarez}\footnote{Unimodular gravity is a speculative approach towards explaining why (the zero mode of) the vacuum energy seems to violate the equivalence principle (the {\em active cosmological constant problem}). The main idea is just to eliminate the direct coupling in the action between the potential energy and the gravitational field. This leads to consider unimodular theories, where  the metric tensor is constrained to be unimodular in the Einstein frame  $g_E\equiv \left|\text{det} g^E_{\m\n}\right|=1.$

The simplest nontrivial such unimodular gravitational action  reads
\bea
&&S_U\equiv -{1\over 16\pi G_n}\int d^n x ~R_E=- M_p^{n-2}\int d^n x~g^{1\over n} \left(~R+{(n-1)(n-2)\over 4 n^2} {g^{\m\n}\nabla_\m g~\nabla_\n g\over g^2}\right)\nonumber
\eea
(the Einstein metric being inert) as well as invariant under area preserving (transverse) diffeomorphisms, that is, those that enjoy unit jacobian, thereby preserving the Lebesgue measure.

} in the gauge
\be
\phi+2^{3\over 2} M_p^{n-2\over 2}\sqrt{n-1\over n-2} g^{-{n-2\over 4 n}}=0
\ee

It is plain that the field redefinition
\be
G_{\m\n}\equiv ~{1\over M_p^2}\left({n-2\over 8(n-1)}\right)^{2\over n-2}~\phi^{4\over n-2}~g_{\m\n}
\ee
transforms the theory back to GR; we are just undoing what we did to get CDG
\be
S=-M_p^{n-2}~\int \sqrt{G}~d^n x~R[G]
\ee

Conformal symmetry with conformal weight $w_\phi={n-2\over 2}$ for the scalar field
\bea
&&\widetilde{\phi}=\Omega^{2-n\over 2}~\phi
\eea
is then  tautological  to the extent that it leaves invariant the metric $G_{\m\n}$. This is non necessarily the case anymore when couplings to matter are considered, because we are going to assume that matter couples to $g_{\m\n}$ instead to $G_{\m\n}$.  Some interesting albeit speculative physical reasons as to why the metric $g_{\m\n}$ could be  the only one physically  observable have been advanced in \cite{Hooft}.
\par
The above considerations are taken as a motivation to  study the non-minimally coupled system gravitational-scalar field in the following sense
\be
S=- \int \sqrt{|g|}d^n x\left(\xi R \phi^2+{1\over 2}(\nabla\phi)^2\right)
\ee

The global sign in front of the action is irrelevant as it stands, but it is the correct one to couple to a matter lagrangian containing matter fields, denoted collectively by $\psi_i~(i=1\ldots N)$
\be
S_{\text{matter}}\equiv\int d(vol) ~L_{\text{matter}}\left(\psi_i,g_{\m\n}\right)
\ee
This sign reflects the gravitational origin of our former lagrangian.
\par
This system has the following property. There is a symmetry
\be
\phi(x)\rightarrow -\phi(x)
\ee
(Which is promoted to an $U(1)$ when the scalar field is complex and $\phi^2$ is replaced by $|\phi|^2$). There are then two different phases, depending on whether the background field vanishes or not. Only the vanishing solution is compatible with the $\mathbb{Z}_2$ symmetry. In the {\em symmetric phase}, we are thus studying quantum perturbations around the symmetric classical solution
\be
\bp(x)=0
\ee

In this case there is no propagator for the gravitational fluctuation, and we do not know how to proceed (athough some possible paths will be suggested in our conclusions). In the {\em broken
phase} we consider a classically nonvanishing solution
\be
\bp(x)\neq 0
\ee
that determines the graviton propagator. Lacking any better option, we shall dub this system {\em dilaton gravity}, although this name is really adequate in the conformal case only; that is, when there is a particular value of $\xi$,
\be
\xi_c\equiv {n-2\over 8(n-1)}
\ee
for which the symmetry is enhanced to full (local) conformal symmetry and we actually recover the CDG mentioned above.
\par

The aim of the present paper is to study dilaton gravity both in the non-conformal as well as in the conformal point. Using a combination of background field and heat kernel techniques, the one-loop effective action will be first determined for generic value of the coupling constant $\xi$. This calculation is not valid at the conformal point, $\xi=\xi_c$, because then there is an enhanced gauge symmetry, namely conformal symmetry. It can be argued that because the classical action of CDG is the group average of Einstein-Hilbert, this should also hold to one loop. Were this true, the counterterms would be derived just by performing a conformal transformation in the 't Hooft -Veltman counterterm  We report a nontrivial one-loop computation in CDG to show that this is indeed the case. In the last section one issue is discussed in some detail, namely the extent to which it is possible to define a renormalized theory which is still Weyl invariant. The conclusion is that in order to do that counterterms with a logarithmic dependence on the dilaton field are needed.

\section{Nonconformal Dilaton Gravity}
Let us begin by analyzing the nonconformal case, that is
\be
S=- \int \sqrt{|g|}d^n x\left(\xi R \Phi^2+{1\over 2}(\nabla\Phi)^2\right)
\ee
where
\be
\xi\neq\xi_c
\ee
The reason for the notation $\Phi$ will be apparent in a moment.
The simplest way to proceed in order to compute the divergences of any action involving the gravitational field is to use heat kernel techniques pioneered by Bryce de Witt. It can be shown \cite{Decanini} that this is equivalent to the assumption that the singular part of the  propagator is of Hadamard type. Those techniques are much less useful to compute finite parts. They are particularly efficient for one loop calculations, which can be reduced to the computation of some determinants, provided the operators in question are minimal ones (otherwise the technique is somewhat unwieldly) There are many reviews avaliable, for example \cite{Barvinsky}. We shall follow a notation similar to \cite{AlvarezF}\cite{Jack}. A brief summary explaining our notation can be found in Appendix \ref{apendixA}.

Let us then expand the action around an arbitrary background
\begin{align}
g_{\m\n}&=\bar{g}_{\m\n}+h_{\m\n}\\
\nonumber \Phi &=\bf+\f
\end{align}
Demanding that the linear terms in the expansion cancel determines the background equations of motion (EM). When the background fields are so restricted,
absence of tadpoles in the quantum theory is guaranteed. In four spacetime dimensions and with arbitrary parameter $\xi$ they read
\begin{align}
\nonumber&\xi \br_{\m\n}=\frac{1}{4}\bg_{\m\n}\frac{\bD^{2}\bf}{\bf}-\left(\frac{1}{2}-2\xi\right)\frac{\bD_{\m}\bf\bD_{\n}\bf}{\bf^{2}}-\left(2\xi-\frac{1}{4}\right)\bg_{\m\n}\frac{(\bD\bf)^{2}}{\bf^{2}}+2\xi\frac{\bD_{\m}\bD_{\n}\bf}{\bf}-2\xi \bg_{\m\n}\frac{\bD^{2}\bf}{\bf}\\
 & \br-\frac{1}{2\xi}\frac{\bD^{2}\bf}{\bf}=0
\end{align}

The result of the expansion of the action to second order in the quantum fields reads
\begin{align}
S_{2}=&-\int d^{n}x\sqrt{|\bg|}\; \left[ h^{\m\n}\hat{H}_{\m\n\r\s}h^{\r\s}+\phi(\widehat{HF})_{\m\n}h^{\m\n}+\phi \hat{F}\phi+\right.\\
&\left.+\xi\bf^{2}\left(-\frac{1}{2}\bD_{\m}h\bD_{\n}h^{\m\n}+\frac{1}{2}\bD_{\m}h^{\m\a}\bD_{\n}h^{\n\a}\right)\right]\nonumber
\end{align}
where we have kept apart the non-diagonal contributions to the graviton sector in order to cancel them later with a proper gauge fixing.
The corresponding second order operators are given in the Appendix \ref{apendixB}.

It is now useful to perform a field redefinition
\be
 k_{\m\n}=\bf h_{\m\n}
 \ee
  in order to eliminate all the dependence on $\bf$ out of the kinetic term. This only makes sense in the broken phase, since this transformation is ill-defined when $\bf=0$. In any other case, the action is thus rewritten as
\begin{align}
S_{2}=&-\int d^{n}x\sqrt{|\bg|}\; \left[k^{\m\n}\hat{H}_{\m\n\r\s}k^{\r\s}+\phi(\widehat{HF})_{\m\n}k^{\m\n}+\phi \hat{F}\phi)+\right.\\
&\left.+\xi\left(-\frac{1}{2}\bD_{\m}k\bD_{\n}k^{\m\n}+\frac{1}{2}\bD_{\m}k^{\m\a}\bD_{\n}k^\n_\a\right)\right]\nonumber
\end{align}
where the explicit values of the coefficients can be found in the Appendix \ref{apendixB}.

The gauge fixing for diffeomorphism (Diff from now on) invariance will be chosen with an eye put on being able to implement heat kernel techiques in the simplest possible way. This indicates that we shall try to  cancel any non-minimal contribution to the kinetic term. In other words, it has to cancel any term in second derivatives which is not proportional to the laplacian, such as the non diagonal terms $\bD_{\m}k^{\m\n}\bD_{\a}k^{\a}_{\n}$ or $\bD_{\n}\f\bD_{\m}k^{\m\n}$.

This can be achieved in different ways, some of them simple modifications of the well-known harmonic or De Donder gauge. It is actually possible  to choose a very general gauge interpolating between two funcions
\begin{align}
\hat{F}_{\m}=(1-\gamma)F^{1}_{\m}+\gamma F^{2}_{\m}
\end{align}
with
\begin{align}
&F_{\m}^{1}=\bD^{\n}k_{\m\n}-\frac{1}{2}\bD_{\m}k -2\bD_{\m}\f\\
&F_{\m}^{2}=\bf\left(\bD^{\n}h_{\m\n}-\frac{1}{2}\bD_{\m}h\right) -2\bD_{\m}\f
\end{align}

Although each of the  two functions $F^{1}_{\m}$ and $F^{2}_{\m}$ represent perfectly admissible gauges separately, we have decided to consider this more general linear combination of them as above in order to be able to track the dependence on the $\gamma$ parameter along the computation and explicitely check that it vanishes  on-shell, as it should.
The full  gauge fixing is then
\begin{align}\label{gaugefixfunction}
\hat{F}_{\m}=\bD^{\n}k_{\m\n}-\frac{1}{2}\bD_{\m}k -2\bD_{\m}\f-\gamma k_{\m}^{\n}\frac{\bD_{\n}\bf}{\bf}+\gamma \frac{1}{2}k\frac{\bD_{\m}\bf}{\bf}
\end{align}
The term to be included in the action then reads
\begin{align}
S_{diff}=\chi \int d^{n}x\sqrt{|\bg|}\; \hat{F}_{\m}\hat{F}^{\m}
\end{align}
with
\begin{align}
&\hat{F}_{\m}\hat{F}^{\m}=\nonumber
2\left(\frac{1}{2}\bD_{\m}k\bD_{\n}k^{\m\n}-\frac{1}{2}\bD_{\m}k^{\m\a}\bD_{\n}k^{\n}_{\a}\right)+2\left[-2\f \bD_{\m}\bD_{\n}k^{\m\n}+\f\bD^{2}k\right]+4\f\bD^{2}\f+\\
\nonumber &+\frac{1}{4}k\bD^{2}k+\gamma\left[k^{\m\n}k_\n^{\a}\frac{\bD_{\m}\bf\bD_{\a}\bf}{\bf^{2}}+\frac{1}{4}k^{2}\frac{(\bD\bf)^{2}}{\bf^{2}}-k k^{\m\n}\frac{\bD_{\m}\bf\bD_{\n}\bf}{\bf^{2}}+k\frac{\bD_{\m}\bf}{\bf}\bD_{\n}k^{\m\n}-\right.\\
 &\left. -2 k^{\m\n}\frac{\bD_{\n}\bf}{\bf}\bD^{\a}k_{\a\m} -\frac{1}{2}k \bD_{\m}k \frac{\bD^{\m}\bf}{\bf}+k^{\m\n}\bD_{\m}k\frac{\bD_{\n}\bf}{\bf}-2k\frac{\bD_{\m}\bf}{\bf}\bD^{\m}\f +4k^{\m\n}\frac{\bD_{\m}\bf}{\bf}\bD_{\n}\f\right]
\end{align}
which cancels exactly the non-minimal terms when $\chi={\xi\over 2}$.

The original action with the gauge fixing added then reads
\begin{align}\label{diff_fixed_action2}
S_2^{full}=-\int d^{n}x\sqrt{|\bg|}\left[k^{\m\n}\hat{H}_{\m\n\r\s}k^{\r\s}+\phi (\widehat{HF})_{\m\n} k^{\m\n}+\phi\hat{F}\phi\right]
\end{align}
where the values of the coefficients are again to be found in the Appendix \ref{apendixB}.
\par
\par

\par
Let us then define a generalized field living in the ''gauge" bundle\footnote{
Capital indices label the different physical fields so that the matrices involved in the action carrying two indices are split in three parts: a $kk$ box carrying four indices (in some sense identifying $A=\m\n$ and $B=\r\s$), another diagonal box corresponding to the $\phi\phi$ element that behaves as a scalar and two non-diagonal blocks carrying two space-time indices $\m\n$ over the diagonal of the matrix and $\r\s$ under it. The rank of the index thus counts the number of physical degrees of freedom (not fields), being $1+ n(n+1)/2$.
}
that includes all the  fields over which we are integrating
\begin{align}
\Psi^{A}=\begin{pmatrix}
k^{\m\n}\\
\phi
\end{pmatrix}
\end{align}

The kinetic term corresponding to  (\ref{diff_fixed_action2}) can then be rewritten as
\begin{align}
-\Psi^{A}G_{A B}\bD^{2}\Psi^{B}
\end{align}
where the metric $G_{AB}$ is symmetric and given by
\begin{align}
G_{AB}=\begin{pmatrix}
 \frac{\xi}{4}\left(\frac{1}{2}{\cal K}_{\m\n\r\s}^{\a\b}-{\cal P}_{\m\n\r\s}^{\a\b}\right)\bg_{\a\b}& \frac{\xi}{2}\bg_{\m\n}\\
\frac{\xi}{2}\bg_{\r\s} & \frac{1}{2}-2\xi
\end{pmatrix}
\end{align}
with inverse
\begin{align}
&G^{AB}=\frac{1}{8 \xi(n-1)-(n-2)}\begin{pmatrix}
 G^{\m\n\r\s}& 8\bg^{\m\n}\\
8\bg^{\r\s}&-2(n-2)
\end{pmatrix}\\
\nonumber &G^{\m\n\r\s}=-\frac{2}{\xi}\left[(8 \xi(n-1)-(n-2))(\bg^{\m\s}\bg^{\n\r}+\bg^{\m\r}+\bg^{\n\s})+2(1-8\xi)\bg^{\m\n}\bg^{\r\s}\right]
\end{align}
defined in such a way that
\begin{align}
G_{AB}G^{BC}=G^{CB}G_{BA}=\begin{pmatrix}
 \frac{1}{2}\left(\delta^{\r}_{\m}\delta_{\n}^{\s}+\delta^{\r}_{\n}\delta^{\s}_{\m}\right)&0\\
0&1
\end{pmatrix}
\end{align}

We will then rewrite the action as
\begin{align}
S_2^{full}=\int d^{n}x\sqrt{|\bg|}\; \Psi^{A}\left(-G_{AB}\bD^{2}+N^{\m}_{AB}\bD_{\m}+M_{AB}\right)\Psi^{B}
\end{align}
whith $N_{AB}^{\m}$ being antisymmetric and $M_{AB}$ being symmetric in their capital indices, meaning interchange of physical field in both sides of the operator, which translates to interchange of the pairs of indices $(\mu\nu)\leftrightarrow (\r\s)$ in the $kk$ elements. Again, the detailed expression of the different matrices are to be found in the Appendix.

\par

To compute the heat kernel coefficient (\ref{heat kernel}) for the previous action, we shall find first the  bundle connection $\omega_\mu$ and the endomorphism $E$ that will allow us to express $S_2^{full}$ as follows
\begin{align}\label{bosact}
S_2^{full}=\int d^{n}x\sqrt{|\bg|}\; \Psi_{A}\left(-\bg^{\mu\nu}[\bD_{\mu} \delta^A_{\phantom{\mu}C}+\omega^{A}_{\mu\,C}][\bD_{\nu} \delta^C_{\phantom{\nu}B}+\omega^{C}_{\nu\,B}]\,-\,E^{A}_{\phantom{\mu}B}\;\right)\Psi^{B},
\end{align}
where $\Psi_{A}=\Psi^{B}G_{BA}$. It can be checked easily that the following equation holds
\begin{align}
G^{AC}\left(-G_{CB}\bD^{2}+N^{\m}_{CB}\bD_{\m}+M_{CB}\right)=-g^{\mu\nu}(\bD_{\mu} \delta^A_{\phantom{\m}C}+\omega^{A}_{\mu\,C})(\bD_{\nu} \delta^C_{\phantom{\mu}B}+\omega^{C}_{\nu\,B})\,-\,\hat{E}^{A}_{\phantom{\m}B},
\end{align}
if
\begin{align}\label{omecon}
&\omega^{A}_{\mu\,B}=\frac{1}{2}G^{AC}\,N_{\mu\,CB}\\
\nonumber&\hat{E}^{A}_{\phantom{\mu}B}=G^{AC}(-M_{CB}- \omega_{\m  C F} \omega^{\m F}_{\phantom{\m}B}-\bD_{\m}\omega_{CB}^{\m})
\end{align}

Now, $\Psi^A\,\bD_{\m}\omega_{AB}^{\m}\Psi^B=0$, for  $\bD_{\m}\omega_{AB}^{\m}$ is antisymmetric under the exchange of $A$ and $B$. Hence, our endomorphism, $E^{A}_{\phantom{\mu}B}$, will be obtained from $\hat{E}^{A}_{\phantom{\mu}B}$ in (\ref{omecon}) by
removing from the latter the contribution   $G^{AC}\bD_{\m}\omega_{CB}^{\m}$, which does not contribute to the dynamics:
\begin{align}\label{endom}
E^{A}_{\phantom{\mu}B}=G^{AC}(-M_{CB}- \omega_{\m  C F} \omega^{\m F}_{\phantom{\m}B})
\end{align}

In summary, it is the coefficient (\ref{heat kernel}) of the heat kernel expansion of the operator
\begin{align}
\Delta = - (\bg^{\mu\nu}[\bD_{\mu} \delta^A_{\phantom{\mu}C}+\omega^{A}_{\mu\,C}][\bD_{\nu} \delta^C_{\phantom{\nu}B}+\omega^{C}_{\nu\,B}]\,+\,E^{A}_{\phantom{\mu}B}),
\end{align}
with $\omega^{A}_{\mu\,B}$ and $E^{A}_{\phantom{\mu}B}$ as given in (\ref{omecon}) and (\ref{endom}), respectively, that will give the pole part of the UV divergent contribution coming from $S^{full}_2$ in (\ref{bosact}). See the Appendix \ref{apendixA} for further information.

In order to finish the computation of the heat kernel coefficient (\ref{heat kernel}), one also needs the field strength $F^{\m\n}_{AB}$ which is worked out by means of the Ricci's identity and has a riemannian part and a bundle part:
\bea
F_{\a\b}\,^A\,_{B}&&=\begin{pmatrix}\frac{1}{2}\left(\br^{\m}\,_{\r}\,_{\a\b}\delta^{\n}_{\s}+\br^{\n}\,_{\r}\,_{\a\b}\delta^{\m}_{\s}+\br^{\m}\,_{\s}\,_{\a\b}\delta^{\n}_{\r}+\br^{\n}\,_{\s}\,_{\a\b}\delta^{\m}_{\r}\right) & \quad\quad 0\\
0 &\quad\quad 0
\end{pmatrix}+\\
&&\quad\quad\quad\bD_{\a}\omega_{\b}\,^{A}\,_{B}-\bD_{\b}\omega_{\a}\,^{A}\,_{B}+\omega_{\a}\,^{A}\,_{C}\omega_{\b}\,^{C}\,_{B}-\omega_{\b}\,^{A}\,_{C}\omega_{\a}\,^{C}\,_{B}\nonumber
\eea

The ghost sector of the theory at hand, which  is a simple subset of the quite involved one needed in the conformal case and discussed in the next section, has the following action
\begin{align}\label{ghostact}
S_{ghost}=\int d^{n}x\sqrt{|\bar{g}|}\; \bar{g}^{\mu\nu}\,\bar{\eta}_\mu s_D\tilde{F}_{\nu},
\end{align}
where $s_D\tilde{F}_{\nu}$ denotes the order-one variation of the gauge-fixing function $\tilde{F}_{\mu}$ in (\ref{gaugefixfunction}) induced by the variations
\bea
&&s_D \bg_{\m\n}=s_D \bp=0\nonumber\\
&& s_D h_{\m\n}={1\over \kappa}\left(\bn_\m\eta_\n+\bn_\n\eta_\m\right)+\eta^\r \bn_\r h_{\m\n}+\bn_\m\eta^\r h_{\r\n}+\bn_\n \eta^\r h_{\r\m}\nonumber\\
&&s_D\phi=\eta^\l \bn_\l\left(\bp+\phi\right)\nonumber
\eea
The symbols $\eta^{\mu}$ and $\bar{\eta}^{\mu}$ denote the ghost and antighost fields, respectively. Of  course, $S_{ghost}$ in (\ref{ghostact}) is obtained from the Faddeev-Poov determinant in the standard fashion.

A little algebra yields the contribution to $S_{ghost}$ that is quadratic in  the quantum fields. This contribution reads
\begin{align}
S^{ghost}_2=\int d^{n}x\sqrt{|\bar{g}|}\; \bar{\eta}^{\rho}\left(-\bar{g}_{\rho\sigma}\bar{\D}^{2}+N^{\mu}_{\phantom{\mu}\rho\sigma}\bar{\D}_{\m}+M_{\rho\sigma}\right)\eta^{\sigma},
\end{align}
where
\begin{align}\label{Nghostnoc}
&N^{\mu}_{\phantom{\mu}\rho\sigma}=-(1-\gamma)\,\bar{g}_{\rho\sigma}\,\frac{\bar{\D}^{\mu}\bp}{\bp}+(1+\gamma)\,\frac{\bar{\D}_{\sigma}\bp}{\bp}\,\delta^{\m}_{\phantom{\m}\rho}+(1-\gamma)\,\frac{\bar{\D}_{\rho}\bp}{\bp}\,\delta^{\m}_{\phantom{\m}\sigma} \\
\nonumber&M_{\rho\sigma}=-\bar{R}_{\rho\sigma}+2\frac{\bar{\D}_{\rho}\bar{\D}_\sigma\bp}{\bp}
\end{align}

The heat kernel coefficient (\ref{heat kernel}) associated to $S^{ghost}_2$ is the corresponding coefficient of the heat kernel expansion of the following  operator
\begin{align}
\Delta^{(ghost)} = - (\bar{g}^{\mu\nu}[\bar{\D}_{\mu} \delta^{\rho}_{\phantom{\rho}\lambda}+\omega^{\phantom{\mu}\rho}_{\mu\phantom{\rho}\lambda}][\bar{\D}_{\nu} \delta^{\lambda}_{\phantom{\lambda}\sigma}+\omega^{\phantom{\nu}\lambda}_{\nu\phantom{\lambda}\sigma}]\,+\,E^{\rho}_{\phantom{\rho}\sigma}),
\end{align}
where
\begin{align}
&\omega^{\phantom{\mu}\rho}_{\mu\phantom{\rho}\lambda}=\bar{g}_{\mu\nu}\omega^{\nu\rho}_{\phantom{\nu\rho}\l},\quad \omega^{\nu\rho}_{\phantom{\nu\rho}\sigma}=-\frac{1}{2} \bar{g}^{\rho\lambda}N^{\mu}_{\phantom{\mu}\lambda\sigma}\\
&E^{\rho}_{\phantom{\rho}\sigma}=-\bar{g}^{\rho\lambda}(M_{\lambda\sigma}+\omega_{\mu\lambda\delta}\omega^{\mu\delta}_{\phantom{\mu\delta}\sigma}+\bar{\D}_{\mu}\omega^{\mu}_{\phantom{\mu}\lambda\sigma})
\end{align}
$N^{\mu}_{\phantom{\mu}\lambda\sigma}$ and $M_{\lambda\sigma}$ are given in (\ref{Nghostnoc}).

Finally, the field strength for the connetion defined by $\bar{\D}_{\mu}+\omega_{\mu}$ runs thus
\begin{align}
F_{\rho\sigma\phantom{\mu}\nu}^{\phantom{\rho\sigma}\mu}=\bar{R}^{\mu}_{\phantom{\mu}\nu\rho\sigma}+\bar{\D}_{\rho}\,\omega^{\phantom{\sigma}\mu}_{\sigma\phantom{\mu}\nu}-\bar{\D}_{\sigma}\,\omega^{\phantom{\rho}\mu}_{\rho\phantom{\mu}\nu}+[\omega_{\rho},\omega_{\sigma}]^{\mu}_{\phantom{\mu}\nu}
\end{align}


Once all the matrices are defined, we can compute the relevant traces both for the bosonic physical fields and for the ghost fields and thus finally write the one-loop (de Witt) effective action   as
\bea
&&\Gamma_{DeW}\left[\bg,\bp\right]=\frac{1}{n-4}\left(A_{2}\left({\textrm bosons}\right)-2A_{2}\left({\textrm ghosts}\right)\right)
\eea
where the ghost sector contributes twice and with a minus sign because the presence of two anticonmuting fields. The final result is
\be
\Gamma_{DeW}\left[\bg,\bp\right]=\frac{1}{n-4}\frac{1}{16\pi^{2}}{1\over g(\xi)}\int d^{4}x~\sqrt{|\bg|}\; a_{2}\left[\bg,\bp\right]
\ee
with
\be
g(\xi)\equiv 720 ~\Ep^2~ \left(2 - 8 \Ep + 4 \left(-1 + 8 \Ep\right)\right)^2
\ee
It is remarkable that the effective action  presents a pole when $\xi=0$, which represents physically a scalar field minimally coupled to the gravitational field. The fact that gravity is dynamical in our case is presumably the reason for this divergence.

To be specific, the gravitational EM in this case read
\be
\bn_\m\bp\bn_\n\bp={1\over 2}\bg_{\m\n}(\bn\bp)^2
\ee
which for $n\neq 2$ imply
\be
\left(\bn\bp\right)^2=0
\ee
In the riemannian case (where the metric is positive definite) means that
\be
\bn_\m\bp=0
\ee
On the other hand, the quadratic gravitational piece of the lagrangian reads
\be
L_h^2\equiv{1\over 2}\sqrt{|\bg|}\left\{\left(h^\m_\a h^{\a\n}-{1\over 2}h h^{\m\n}\right)\bn_\m\bp\bn_\n\bp+{1\over 4}\left(\bn\bp\right)^2\left({1\over 2}h^2- h_{\a\b}h^{\a\b}\right)\right\}
\ee
The fact that  it can be written without any derivative acting on the gravitational quantum fluctuations means that the corresponding high frequancy modes are generically not suppressed.
\par
The scalar quadratic piece on the other hand is perfectly kosher
\be
L_s^2\equiv{1\over 2}\sqrt{|\bg|}\bg^{\m\n}\bn_\m\phi\bn_\n\phi
\ee

\par
The heat kernel coefficient $a_{2}$ reads
\small
\bea
&&a_{2}\left[\bg,\bp\right]=(12\xi -1)\left(P_0(\xi,\g)\frac{\D^{\m}\bf\D^{\n}\bf\D_{\m}\bf\D_{\n}\bf}{\bf^{4}}P_1(\xi,\g)\frac{\bD^{\a}\bf\bD^{\b}\bf\bD_{\a}\bD_{\b}\bf}{\bf^{2}}+P_2(\xi,\g)\frac{\bD_{\m}\bD_{\n}\bf\bD^{\m}\bD^{\n}\bf}{\bf^{2}}+\right.\nonumber\\
&&\left.+P_3(\xi,\g)\frac{(\bD\bf)^{2}\bD^{2}\bf}{\bf^{3}} \right)+P_4(\xi,\g)\frac{\bD^{2}\bf\bD^{2}\bf}{\bf^{2}}+P_5(\xi,g)\frac{\bD^{\a}\bD^{\b}\bf\;\br_{\a\b}}{\bf}+P_6(\xi,\g)\frac{\bD^{\a}\bf\bD^{\b}\bf\;\br_{\a\b}}{\bf^{2}}-\nonumber\\
&&-P_7(\xi,\g)\br_{\m\n}\br^{\m\n} +P_8(\xi,\g)\frac{(\bD\bf)^{2}\br}{\bf^{2}}+P_9(\xi,\g)\frac{\bD^{2}\bf\;\br}{\bf}+P_{10}(\xi,\g)\br^{2}+ P_{11}(\xi,\g) \br_{\m\n\a\b}\br^{\m\n\a\b}\nonumber\\
&&
\eea
\normalsize
where the polynomials $P_i(\xi,\g)$ are defined by
\small
\bea
&&P_0(\xi,\g)\equiv 720(-5 + 104 \Ep - 728 \Ep^2 + 2784 \Ep^3 - 18 \Ep \gamma + 72 \Ep^2 \gamma + 1536 \Ep^3 \gamma + 8 \Ep \gamma^2 -  260 \Ep^2 \gamma^2 + 2064 \Ep^3 \gamma^2 -\nonumber \\
&&-16 \Ep^2 \gamma^3 + 216 \Ep^3 \gamma^3)\nonumber\\
&&P_1(\xi,\g)\equiv -960\xi\left(-29+450\xi-840\xi^2-15\g+88\xi\g+912\xi^2\g-38\xi\g^2+508\xi^2\g^2 -8\xi\g^3+108\xi^2\g^3\right)\nonumber\\
&&P_2(\xi,\g)\equiv 480 \Ep  \left(1 - 78 \Ep + 984 \Ep^2 - 68 \Ep \gamma +  720 \Ep^2 \gamma - 2 \Ep \gamma^2 + 28 \Ep^2 \gamma^2\right)\nonumber\\
&&P_3(\xi,\g)\equiv-480 \Ep \left(-2 + 228 \Ep - 3072 \Ep^2 + 9 \gamma + 64 \Ep \gamma -1680 \Ep^2 \gamma + 16 \Ep \gamma^2 - 368 \Ep^2 \gamma^2 -
   8 \Ep \gamma^3 + 108 \Ep^2 \gamma^3\right) \nonumber\\
&&P_4(\xi,\g)\equiv -480 \Ep \left(-1 - 48 \Ep + 672 \Ep^2 - 3312 \Ep^3 +56 \Ep \gamma - 1248 \Ep^2 \gamma + 6912 \Ep^3 \gamma + 2 \Ep \gamma^2 -
   52 \Ep^2 \gamma^2 + 336 \Ep^3 \gamma^2\right)\nonumber\\
&&P_5(\xi,\g)\equiv -3840 \Ep^2 \left(-1 + 12 \Ep) (3 - 12 \Ep -\gamma+ 6 \Ep \gamma\right)\nonumber\\
&&P_6(\xi,\g)\equiv -480 \Ep (-1 + 12 \Ep) \left(-1 + 42 \Ep -744 \Ep^2 + 52 \Ep \gamma -
   528 \Ep^2 \gamma - 10 \Ep \gamma^2 + 116 \Ep^2 \gamma^2\right)\nonumber\\
&&P_7(\xi,\g)\equiv -48 \Ep^2 (-1 + 12 \Ep) \left(-241 + 2412 \Ep\right)\nonumber\\
&&P_8(\xi,\g)\equiv -960 \Ep (-1 + 12 \Ep) \left(1 - 41 \Ep + 432 \Ep^2 - 32 \Ep \gamma + 348 \Ep^2 \gamma - 6 \Ep \gamma^2 +90 \Ep^2 \gamma^2\right)\nonumber\\
&&P_9(\xi,\g)\equiv 1920 \Ep^2 \left(-11 + 189 \Ep - 1008 \Ep^2 + \gamma - 18 \Ep \gamma +
   72 \Ep^2 \gamma\right)\nonumber\\
&&P_{10}(\xi,\g)\equiv 120 \Ep^2 \left(29 - 576 \Ep +3168 \Ep^2\right)\nonumber\\
&&P_{11}(\xi,\g)\equiv 3408 \Ep^2 (-1 + 12 \Ep)^2
\eea
\normalsize
There is a set of different terms appearing in the counterterm that will be related both by the EM as well as by integration by parts. It is a fact that there are  only three linearly independent monomials. The full set of monomials compatible with the symmetries and dimensional counting which appear in the counterterm is
\begin{align*}
\begin{array}{ll}
G_1\equiv \frac{\bD_{\m}\bf\bD_{\n}\bf\; \br^{\m\n}}{\bf^{2}}\quad\quad\quad
&A=\frac{\bD^{2}\bf\bD^{2}\bf}{\bf^{2}}\\
\\
G_2\equiv \frac{\bD_{\m}\bD_{\n}\bf\; \br^{\m\n}}{\bf}\quad\quad\quad
& B=\frac{\bD^{2}\bf (\bD\bf)^{2}}{\bf^{3}}     \\
\\
G_3\equiv \frac{\bD^{2}\bf\;\br}{\bf}\quad\quad\quad
&
C=\frac{(\bD\bf)^{2}(\bD\bf)^{2}}{\bf^{4}} \\
\\
G_4\equiv \frac{(\bD\bf)^{2}\br}{\bf^{2}}\quad\quad\quad
&D=\frac{\bD_{\m}\bD_{\n}\bf\bD^{\m}\bD^{\n}\bf}{\bf^{2}} \\
\\
G_5\equiv \br_{\m\n}\br^{\m\n}\quad\quad\quad
&E=\frac{\bD_{\m}\bf\bD^{\n}\bf\bD^{\m}\bD_{\n}\bf}{\bf^{3}}\\\
\\
G_6\equiv \br^{2}\quad\quad\quad
&F=\frac{\bD_{\m}\bf\bD^{2}\bD^{\m}\bf}{\bf^{2}}\\
\\
G_7\equiv \br_{\m\n\a\b}\br^{\m\n\a\b}&
\end{array}
\end{align*}

The EM impose some relations between them, namely
\begin{align*}
&G_3\equiv \frac{\bD^{2}\bf\;\br}{\bf}=\frac{1}{2\xi}A\\
&G_4\equiv \frac{(\bD\bf)^{2}\br}{\bf^{2}}=\frac{1}{2\xi}B\\
&G_6\equiv \br^{2}=\frac{1}{4\xi^{2}}A\\
&G_1\equiv \frac{\bD^{\m}\bf\bD^{\n}\bf\;\br_{\m\n}}{\bf^{2}}=\left(\frac{1}{4\xi}-2\right)B-\frac{1}{4\xi}C+2E\\
&G_2\equiv \frac{\bD_{\m}\bD_{\n}\bf\; \br^{\m\n}}{\bf}=\left(\frac{1}{4\xi}-2\right)(A+B)+2D+\left(2-\frac{1}{2\xi}\right)E\\
&G_5\equiv \br_{\m\n}\br^{\m\n}=\left(\frac{1}{4\xi}-2\right)\br\left[\frac{\bD^{2}\bf}{\bf}+\frac{(\bD\bf)^{2}}{\bf^{2}}\right]+\left(2-\frac{1}{2\xi}\right)\frac{\bD^{\m}\bf\bD^{\n}\bf\;\br_{\m\n}}{\bf^{2}}+2\frac{\bD^{\m}\bD^{\n}\bf\; \br_{\m\n}}{\bf}
\end{align*}
and by using this and integrating by parts, it can be shown that $D$, $E$ and $F$ can be written in terms of $A$, $B$ and $C$,
\begin{align*}
&\int d(vol) D=\int d(vol)\left( A-2B+2E-\frac{\bD^{\m}\bf\bD^{\n}\bf\;\br_{\m\n}}{\bf^{2}}\right)\\
&\int d(vol) E=\int d(vol)\left(\frac{3}{2} C -\frac{1}{2} B\right)\\
&\int d(vol)  F=\int d(vol)  \left(-D+2E\right)
\end{align*}

\par
Finally, whenever $\xi\neq \frac{1}{12}$ there is an extra relation that we can use and that comes from the fact that the two equations of motion for the metric and the scalar field must be compatible. Taking the trace of  the first one we have
\begin{align}
\br=\left(\frac{n}{4\xi}+2-2n\right)\frac{\bD^{2}\bf}{\bf}+\left(\frac{n}{4\xi}+2-2n-\frac{1}{2\xi}\right)\frac{(\bD\bf)^{2}}{\bf^{2}}
\end{align}
so requiring  agreement with the scalar equation of motion requires
\begin{align}
\frac{\bD^{2}\bf}{\bf}+\frac{(\bD\bf)^{2}}{\bf^{2}}=0
\end{align}
which implies
\be
A=C=-B
\ee

In the case $\xi=\frac{1}{12}$ this identity is satisfied identically and these last relations cannot be used.

\par
When the background fields are put on-shell and  the preceding identities are taken into account,  all the dependence in the gauge fixing parameter $\gamma$ dissapears (This is just DeWitt-Kallosh' theorem; cf. also \cite{Grisaru}) and we end up with
\begin{align}
\left.\Gamma_{DeW}\right|_{\text{on shell}}=\frac{1}{n-4}\frac{1}{16\pi^{2}}\int d^{4}x\sqrt{|\bg|}\; \left(\frac{71}{60}~W_4+\frac{1259}{1440}\frac{(1 - 12 \Ep)^2}{\xi^{2}}\frac{(\bD\bf)^{2}(\bD\bf)^{2}}{\bf^{4}}\right)
\end{align}

The  Euler density (the quantity whose integral yields the Euler characteristic) is given by
\be
E_4\equiv \br_{\m\n\r\s}\br^{\m\n\r\s}-4 \br_{\m\n}\br^{\m\n}+\br^2
\ee

It is a fact that
\be
~W_4=~2\left(\br_{\m\n}^2-{1\over 3}~\br^2\right)+~E_4
\ee

This means that on Einstein-Hilbert's shell (that is, when spacetime is Ricci-flat) $E_4$ and $W_4$ are equivalent.
  When the space is Ricci-flat and Euler's characteristic vanishes,  then
  \be
  \int d(vol)~\br_{\m\n\r\s}^2=0
  \ee
   as well.
Usually the anomaly integrand is represented as
\be
a~E_4 - c ~W_4
\ee
which reduces on Einstein-Hilbert  shell to
\be
\left(a-c\right)~E_4
\ee

In the present situation we can assert that
\be
\int d(vol)~\left(E_4-W_4\right)=\left(-1+12\xi\right)\int d(vol){A+3C + 48 \xi\left( B-C\right)\over 12 \xi^2}
\ee

It is worth stressing that the value of this coefficient is different from the one that we will find in the conformal case, when $\xi=\xi_c$.

\section{ Conformal Ward identities.}
Let us now shift to the conformal case, id est,
\be
\xi=\xi_c
\ee
The framework is then a theory including the metric as well as a set of matter fields, $\psi_i$, with scale dimensions $\l_i$, which is conformal.
\par
Let us now spell out the consequences of conformal symmetry at the quantum level.
We can start with the path integral with external sources in it
\bea
Z\left[J^{\m\n},J\right] \equiv \int {\cal D} g_{\m\n}~  {\cal D}\phi ~e^{i S[g_{\m\n}\phi]+i\int d(vol)\left( J^{\m\n} g_{\m\n }+ J ~\phi\right)}
\eea

The gravitational equations of motion (EM) read
\be
\int {\cal D} g_{\m\n}~  {\cal D}\phi~{1\over i}~{\d\over \d g^{\m\n}} ~e^{i S[g_{\m\n}\phi]+i\int d(vol)\left( J^{\m\n} g_{\m\n }+ J ~\phi\right)}\equiv \left\langle 0_+\left|{\d S\over \d g_{\m\n}}+J^{\m\n} \right|0_-\right\rangle=0
\ee

Those are operator equations (id est, their expectation values between any pair of states vanishes). It is obvious that if the EM are valid then of course its trace (which is a linear combination of EM) also vanishes; in the absence of sources,
\be
g^{\m\n}\left\langle 0_+\left|{\d S\over \d g^{\m\n}} \right|0_-\right\rangle=0
\ee

It is worth emphasizing that this is a much subtler concept that the tracelessness of the energy-momentum tensor in a conformal quantum field theory in an external gravitational field. The energy-momentum tensor does not vanish; it is only covariantly conserved, and this does not imply tracelessness even on shell.
\par
Here if we want the statement to have any content, what is implied is that the trace of the EM is not itself an EM, because it vanishes identically {\em without the use of the EM}. Let us consider the path integral defining the partition function. We could as well perform the path integral using Weyl-transformed variables. They are dummies, after all. Demanding that the difference between the two different ways of computing the integral should vanish leads to the whole hierarchy of conformal Ward identities. Let us write them down for CDG (where $\d g_{\m\n}=2\omega(x) g_{\m\n}$)
\bea
&&0=\d Z\equiv \int {\cal D} g_{\m\n}~ \prod_i {\cal D}\psi_i ~\int d(vol)_x \omega(x)\bigg\{-2 g^{\m\n}(x){\d S\over \d g^{\m\n}(x)}-{n-2\over 2}\phi{\d S\over \d \phi}+\nonumber\\
&&+2 J^{\m\n}(x) g_{\m\n}(x)- J(x) \phi(x)\bigg\}~\exp~\bigg\{i S[g_{\m\n}\phi]+\int d(vol)\left( J^{\m\n} g_{\m\n }+ J ~\phi\right)\bigg\}
\eea

When the sources vanish, this conveys the fact that the equations of motion must be traceless not only classically as a Noether identity, but also its expectation value between any pair of states that are connected through the path integral with appropiate boundary conditions. The vacuum expectation value is a particular case of it when all sources are switched off.
\ben\label{tracelessness}
\left\langle 0_+\left|g^{\m\n}(x){\d S\over \d g^{\m\n}(x)}+{n-2\over 4}\phi{\d S\over \d\phi}\right|0_-\right\rangle=0
\ee

We emphasize that those identities are true {\em off shell}; that is without the use of the EM.
Taking derivatives with respect to the sources yield all contact terms that appear in higher correlators.
\par

It is convenient at this stage to reflect on this result. The equation of motion for the graviton is proportional to the energy-momentum tensor the graviscalar field would had if gravitation were not dymamical.
\be
{\d S^{CDG}\over \d g^{\m\n}}={2\over \sqrt{|g|}} T_{\m\n}
\ee
The Ward identity then tells us that when gravitation becomes dynamical, the trace of the energy-momentum tensor is {\em off-shell} proportional to the equation of motion of the graviscalar.
\be
\left\langle 0_+ \left|g^{\m\n}T_{\m\n}\right|0_-\right\rangle=-{n-2\over 4}~\left\langle 0_+\left|{1\over \sqrt{|g|}}\phi{\d S^{CDG}\over \d \phi}\right|0_-\right\rangle
 \ee
{\em On shell} both terms vanish trivially.
\par
What characterizes conformal invariant theories with dynamical gravity is precisely this conformal Ward identity. We shall investigate in due time whether the effective action still fulfills it after taking loop contributions into account. A technical problem is the following. The effective action (which coincides with the background field free energy at one loop) is gauge dependent off shell. When we restrict to on shell quantities, the Ward identity as such looks trivial (because it is a linear combination of the expectation values of the equations of motion). It is well-known however \cite{Duff} that when there are evanescent operators in the divergent part, id est, operators such that
\be
\d E[\phi]\sim (n-4) E[\phi],
\ee
then the Ward identity expressing conformal invariance is violated.
\par
In terms of the singlet metric $G_{\a\b}$ the classical EM  read
\be
R_{\m\n}[G]=0=R_{\m\n}+{2 n \over n-2}{\nabla_\m\phi\nabla_\n\phi\over\phi^2}-2{\nabla_\m\nabla_\n\phi\over\phi}-{2\over n-2}\left({(\nabla\phi)^2\over\phi^2}+{\nabla^2\phi\over\phi}\right)g_{\m\n}
\ee

When varying the two fields in the CDG in an independent way, the EM read
\bea
&&{\d S^{CDG}\over \d \phi}\equiv -\nabla^2 \phi+{n-2\over 4(n-1)}~R~\phi=0\\
&&{8(n-1)\over n-2}{\d S^{CDG}\over \d g^{\a\b}}\equiv R_{\a\b}~\phi^2+{2 n\over n-2} \nabla_\a\phi\nabla_\b\phi- 2 \phi\nabla_\a\nabla_\b\phi-\nonumber\\
&&-{1\over 2}\left(R \phi^2+{4\over n-2}\left(\nabla\phi\right)^2-4 \phi\nabla^2\phi \right)g_{\a\b}=0
\eea

It is then a fact that at tree level
\be
2 g^{\m\n}{\d S^{CDG}\over \d g^{\m\n}}+{n-2\over 2}~\phi{\d S^{CDG}\over \d \phi}\equiv -{\d S\over \d w(x)}=0
\ee

This is a fundamental identity which carries several consequences. First of all, it means that the two set of EM are  compatible at the classical level.
\par
 But it also embodies the Noether identity
 \be
 {\d S\over \d w(x)}\equiv 0
 \ee
  associated to the conformal invariance of the action. The Weyl transformation of the metric is compensated by a conformal transformation of the scalar. The corresponding Ward identy on the effective action implies that the possible conformal anomaly in the gravitational sector should be cancelled by the contribution of the gravitational scalar. One of the main objectives of the present paper is to examine whether this is the case.
\par
Actually, in the present paper we shall confine ourselves to pure CDG in the absence of any matter.

\par
In order to integrate over the gravitational fluctuations, it  is much simpler to work with the singlet metric $G_{\m\n}$. Let us be specific. Given the fact that {\em classically} CDG is nothing but the group averaged action of Einstein-Hilbert under conformal transformations, we could {\em conjecture} that the same is true in the quantum theory; that is, that the counterterm of CDG can be obtained from the 't Hofft-Veltman one by the group-averaging procedure. This conjecture needs of course an explicit verification before it is accepted. We shall do such a calculation in the next paragraph.
\par

At any rate, there is an infinite factor coming from the functional integration over the gravitational scalar, which does not appear in the action. This infinite factor disappears in all connected amplitudes. We are {\em defining}
\bea
&& e^{i \Gamma\left[\bar{g}_{\m\n}, \bar{\phi}_g\right]}\equiv \int {\cal D} g_{\m\n}~{\cal D}\phi~e^{-i {1\over 2}\int d^4 x~\sqrt{-g}\left(\pd_\m\phi ~\pd^\m\phi+{1\over 6}~R~\phi^2\right)}
\eea
through
\be
e^{i \Gamma\left[\bar{g}_{\m\n}, \bar{\phi}_g\right]}:=e^{i \Gamma\left[\bar{G}_{\m\n}\left[\bar{g}_{\m\n},\bar{\phi}_g\right]\right]}
\ee
where
\bea
&& e^{i \Gamma\left[\bar{G}_{\m\n}\right]}\equiv \int {\cal D} G_{\m\n}~e^{{i\over 16\pi G} \int d^4 x R\left[G_{\m\n}\right]}
\eea
and the function
\be
\bar{G}_{\m\n}\left[\bar{g}_{\m\n},\bar{\phi}_g\right]\equiv {1\over M_p^2}~\left({n-2\over 8(n-1)}\right)^{2\over n-2}~\bp^{4\over n-2}~\bg_{\m\n}
\ee

Actually there is in the best of cases a divergent proportionality factor, so that the equivalence is as best true for the connected piece, which we precisely denote the effective action, $W$. In the particular case of the Einstein-Hilbert term, the effective action is nothing but the well-known 't Hooft-Veltman \cite{'tHooftV} \cite{Barvinsky}counterterm for pure gravity. This yields

\bea
&& \Gamma_{DeW}\left[\overline{G}\right]=\frac{1}{\pi^{2}(n-4)}\int d^4 x\sqrt{|\overline{G}|}\left({149\over 2880}E_4[\overline{G}]+{7\over 320} W_4[\overline{G}]+{3\over 128}R[\overline{G}]^2\right)
\eea

Given the fact that the integral of the Weyl tensor squared is conformally invariant, we can naively put $G\rightarrow g$ on that term. If we keep the spacetime dimension at the generic value, the result is
\be
\int d(vol)~W_4\left[\Omega^2 g_{\m\n}\right]=~\int d(vol)~\Omega^{n-4}~W_4\left[g_{\m\n}\right]
\ee

This is due to the fact that the covariant Weyl tensor has conformal weight $-2$ in {\em any dimension}, whereas the volume element picks a factor $\Omega^n$. The same thing happens with the integral of the Euler density
 \be
\int d(vol)~E_4\left[\Omega^2 g_{\m\n}\right]=~\int d(vol)~\Omega^{n-4}~E_4\left[g_{\m\n}\right]
\ee

The term in $R^2$ is not conformal invariant in any dimension.
\par
The variation of the action under a conformal transformation is then an {\em evanescent operator}. This means simply that it is proportional to $(n-4)$. By itself, it vanishes when $n\rightarrow 4$, but when (as is here the case) is multiplied by a pole term, it yields a finite contribution.
This has in turn the important consequence that the one loop expectation value of the {\em trace} of the equations of motion (this is the analogous to the energy-momentum tensor when gravity is dynamical) does not vanish
\be
\left\langle g^{\m\n}~{\d S\over \d g^{\m\n}}\right\rangle=2 \left.{\d S_{\text{eff}}\over\d\Omega}\right|_{\Omega=1}\neq 0
\ee

This is the analogous of the {\em conformal anomaly} and we shall dub it as such.
\par

The total result for the divergent piece in four dimensions assuming the hypothesis as above is then
\bea
&& \Gamma_{DeW}\left[\bp,\bg\right]=\frac{1}{\pi^{2}(n-4)}\int d(vol_{\bg}) \bigg\{{149\over 2880}E_4[\bg]+{7\over 320} W_4[\bg]+{3\over 128}\left(R[\bg]-6 {\nabla^2\bp\over\bp}\right)^2\bigg\}\nonumber\\
&&
\eea

The piece involving the gravitational scalar also yields a conformal anomaly, because the general formula
\be
\left(\tilde{\nabla}^2-{n-2\over 4(n-1)}\tilde{R}\right)\left(\Omega^{-{n-2\over 2}}\phi\right)=\Omega^{-{n+2\over 2}}\left(\nabla^2-{n-2\over 4(n-1)} R\right)
\ee
implies that
\be
\left(\tilde{R}- {4(n-1)\over n-2}{\tilde{\nabla}^2\tilde{\phi}_g\over \tilde{\phi}_g}\right)^2=\Omega^{-4}~\left(R- {4(n-1)\over n-2}{\nabla^2\phi\over \phi}\right)^2
\ee
which yields again a factor of $\Omega^{n-4}$ when combined with the  n-dimensional riemannian measure. The anomalous Ward identity of the four dimensional CDG then reads
\be
\left\langle 0_+\left|-2 g^{\m\n}{\d S_{CDG}\over \d g^{\m\n}}-{n-2\over 2}\phi {\d S_{CDG}\over \d \phi}\right|0_-\right\rangle\equiv A_{CDG}=\frac{1}{\pi^{2}} \bigg\{{7\over 320} W_4+{3\over 128}\left(R-6 {\nabla^2\phi\over\phi}\right)^2\bigg\}
\ee

The expression of the anomaly is manifestly  pointwise conformally invariant.
It is interesting to compare this result with the cohomological analysis of Bonora, Cotta-Ramusino and Reina \cite{Bonora}. They admit only polynomial candidates for the cocycles.
The  cocycles which are not exact are
\bea
&& C_1\equiv W_4\nonumber\\
&&C_2\equiv E_4\nonumber\\
&&C_3\equiv \phi\nabla^2\phi-{1\over 6}R\phi^2\nonumber\\
&&C_4\equiv \phi^4
\eea

Our expression for the anomaly is clearly of the form
\be
a C_1+ b {C_3^2\over C_4}
\ee
with a and  b constants.

\newpage

\section{Conformal Dilaton Gravity}

Let us now consider the theory at the conformal point, corresponding to the critical coupling $\xi=\frac{(n-2)}{8(n-1)}$. In this case, we have an enhanced gauge symmetry and, as a consequence, an extra zero mode that will make the metric $G_{AB}$ non-invertible, as  can easily be verified  from the fact that its determinant when $\xi=\xi_{c}$ is
\begin{align}
Det(G_{AB})=\frac{n}{64(n-1)}Det\left(\bg_{\m\r}\bg_{\n\s}+\bg_{\m\s}\bg_{\n\r}-\frac{2}{n}\bg_{\m\n}\bg_{\r\d}\right)
\end{align}
which is the determinant of a projector.
\par

In this section, by using  the background field method in conjunction with the BRST formalism, we shall quantize the CDG  with  classical action
\be
S_{CDG}=\int d(vol)~\left(-{n-2\over 8(n-1)}~R~ \Phi^2-{1\over 2}g^{\m\n}\nabla_\m\Phi\nabla_\n \Phi\right)
\ee
around a classical field configuration $(\bar{g}_{\mu\nu},\bar{\phi})$ with $\bp\neq 0$. Thus we shall  split first the full fields, $g_{\mu\nu}$  and $\Phi$, entering $S_{CDG}$ into their backgrounds, $\bar{g}_{\mu\nu}$ and $\bar{\phi}$,  and their quantum, $h_{\mu\nu}$ and $\phi$, parts, respectively:
\bea
&&g_{\m\n}\equiv \bg_{\m\n}+ \kappa h_{\m\n}\nonumber\\
&&\Phi\equiv \bp+\phi.
\eea

Then we shall translate the invariance of $S_{CDG}$ under diffeomorphisms and Weyl transformations into its invariance  under the following infinitesimal quantum gauge transformations
\bea
&&\d^Q_D \bg_{\m\n}=\d^Q_D \bp=\d^Q_W \bg_{\m\n }=\d^Q_W\bp=0\nonumber\\
&&\d^Q_D  h_{\m\n}=\xi^\r\bn_\r h_{\m\n}+\bn_\m\xi^\r h_{\r\n}+\bn_\n\xi^\r h_{\m\r}+{1\over \kappa}\left(\bn_\m\xi_\n+\bn_\n\xi_\m\right)\nonumber\\
&&\d^Q_W  h_{\m\n}=2 \omega \left(h_{\m\n}+{1\over \kappa}\bg_{\m\n}\right)\nonumber\\
&&\d^Q_D \phi=\xi^\r\bn_\r\left(\bp+\phi\right)\nonumber\\
&&\d^Q_W \phi=-{n-2\over 2}~\omega\left(\bp+\phi\right).
\eea
The subscripts $D$ and $W$ remind us that the corresponding transformations  either come from diffeomorphisms --$D$-- or from Weyl transformations --$W$.

Since along the quantization process we shall have to handle two different gauge symmetries as the same time, the task of quantization may appear to be slightly tricky. And yet, we shall see below that the BRST quantization method does
the job for us easily. So, next, we shall introduce the  BRST operators, $s_D$ and $s_W$, associated to the previous infinitesimal quantum gauge transformations. These operators act on the fields $\bg_{\m\n}$, $\bp$, $h_{\mu\nu}$ and $\phi$  as follows
\bea
&&s_D \bg_{\m\n}=s_D \bp=s_W\bg_{\m\n }=s_W \bp=0\nonumber\\
&& s_D h_{\m\n}={1\over \kappa}\left(\bn_\m\eta_\n+\bn_\n\eta_\m\right)+\eta^\r \bn_\r h_{\m\n}+\bn_\m\eta^\r h_{\r\n}+\bn_\n \eta^\r h_{\r\m}\nonumber\\
&&s_W h_{\m\n}={2c\over \kappa}\left(\bg_{\m\n}+\kappa h_{\m\n}\right)\nonumber\\
&&s_D\phi=\eta^\l \bn_\l\left(\bp+\phi\right)\nonumber\\
&&s_W\phi=-{n-2\over 2} c\left(\bp+\phi\right)\nonumber\\
\eea

The symbols $\eta^{\mu}$ and $c$ denote the ghost fields for diffeomorphisms and Weyl transformations, respectively. The action of $s_D$ and $s_W$ on $\eta^{\mu}$ and $c$ is given by
\bea
&&s_D\eta^\m=\eta^\r\partial_\r \eta^\m\nonumber\\
&&s_W\eta^\m=0\nonumber\\
&&s_D c=\eta^\r\partial_\r c\nonumber\\
&&s_W c=0.
\eea

To construct a gauge-fixing term that is BRST exact, we shall need the antighost fields, $\bar{\eta}^\m$ and $\bar{c}$, and the corresponding Nakanishi-Lautrup auxiliary fields, $B^{\mu}$ and $f$. The BRST variations of these fields read
\bea
&&s_D\bar{\eta}^\m= B^\m \quad  s_D B^\m=0 \nonumber\\
&& s_W \bar{\eta}^\m=0 \quad s_W B^\m=0\nonumber\\
&&s_D\bar{c}=\eta^\r\partial_\r \bar{c}\quad s_D f=\eta^\l\partial_\l f\nonumber\\
&&s_W \bar{c}= f\quad s_W f=0.
\eea

It can be shown that
\be
s^2_D=0\quad s_W^2=0\quad \left\{s_W,s_D\right\}=0;
\ee
and hence one can introduce the following BRST operator
\be
s=s_D+s_W,
\ee
which takes care at once of both the BRST symmetry associated to diffeomorphisms and the BRST symmetry stemming from Weyl transformations. Clearly, $s^2=0$.

We are now ready to introduce the action $S$ of the BRST quantized theory:
\ben\label{quantumaction}
S= S_{CDG}+ s\,\left(X_D+X_W\right),
\een
where
\bean\label{XDW}
&&X_D= \int d^n x \sqrt{|\bar{g}|~}\;\,\bar{\eta}_\mu\left(-\frac{4(n-1)}{n-2} B^\mu+F_D^{\mu}\right)\\
&&X_W= \int d^n x \sqrt{|g|} ~g^{\m\n}~\partial_\m\bar{c}\,\partial_\n \left(f-\alpha\Phi\right)+\alpha \int d^n x \sqrt{|\bar{g}|} ~\bar{g}^{\m\n}~\partial_\m\bar{c}\,\partial_\n\bp,
\eean
and
\be
F_D^\nu=(1-\gamma)\left(\bn^\m k_{\m}^{\nu}-{1\over 2}\bn^\n k\right)+\gamma\, \bar{\phi}\left(\bn^\m h_{\m}^\n-{1\over 2}\bn^\n h\right) -2\bn^\n\phi.
\ee
Here, and in the sequel, $k_{\mu\nu}=\bp\, h_{\mu\nu}$. $\alpha$ and $\gamma$ are gauge parameters.

Furnished with $S$ as given in the previous equation, we define the DeWitt effective action, $\Gamma_{DeW}[\bar{g}_{\mu\nu},\bar{\phi}]$, of the theory as follows
\bea
&&e^{i\Gamma_{DeW}[\bar{g}_{\mu\nu},\bar{\phi}]}=\\
&&=\int\!{\cal D}h_{\mu\nu}\,{\cal D}\phi\, {\cal D}\eta^{\mu}\,{\cal D}\bar{\eta}^{\mu}\,{\cal D}B^{\mu}\,{\cal D}c\,{\cal D}\bar{c}\, {\cal D}f\;e^{i\left[S-\int d^n x \left(\frac{\delta S_{CDG}[0,0]}{\delta \bar{g}_{\mu\nu}(x)}h_{\mu\nu}(x)+\frac{\delta S_{CDG}[0,0]}{\delta \bp(x)}\phi(x)\right)\right]}\nonumber
\eea
where $S_{CDG}[0,0]$ is obtained by setting $h_{\mu\nu}=0$ and $\phi=0$ in $S_{CDG}[\bar{g}_{\mu\nu}+\kappa h_{\mu\nu},\bar{\phi}+\phi]$.

Taking advantage of the fact that $s\,\left(X_D+X_W\right)$ is BRST exact,  one can show that the appropriately regularized --eg, by using dimensional regularization-- $\Gamma_{DeW}[\bar{g}_{\mu\nu},\bar{\phi}]$ does not depend on the choice of $X_D+X_W$, if $\bar{g}_{\mu\nu}$ and $\bp$ are on-shell. Our choice of $X_D+X_w$ is dictated, partially, by the requirement of having a one-loop contribution to  $\Gamma_{DeW}[\bar{g}_{\mu\nu},\bar{\phi}]$ that is given by a minimal operator. Indeed, on the one hand, as we shall see below,  the contribution to
$S_{CDG}$ which is quadratic on the quantum fields contains a non-minimal part which reads
\be
-\frac{n-2}{16(n-1)}\int\!d x\,\sqrt{|\bar{g}|}\,\left(\bn_{\mu}k^{\mu\nu}\bn^{\lambda}k_{\lambda\nu}-\bn_{\mu}k^{\mu\nu}\bn_{\n}k+4\bn_{\mu}\bn_\nu k^{\mu\nu}\phi\right).
\ee

The need to cancel this term pins down the contribution to $ F_D^\nu$ which does not depend on the gauge parameter $\gamma$. On the other hand, the fact that one can define the action of $s_D$ on $\bar{c}$ and $f$ in such a way that the
result is geometrical makes it possible to construct easily a contribution to $X_W$ that is non-linear in the quantum fields and is annihilated by $s_D$; this contribution being
\ben\label{twoderiv}
\int d^n x \sqrt{|g|} ~g^{\m\n}~\partial_\m\bar{c}\,\partial_\n \left(f-\alpha\Phi\right).
\een
Notice that in the previous expression it is $\Phi$ --the full scalar field-- that occurs, not just $\phi$.

Now, the action of $s_D$ on  $h_{\mu\nu}$ and $\phi$ gives rise to a derivative of the appropriate quantum field.  Hence the $s_D$ variation of terms which --like the previous one-- contain two derivatives of the quantum fields, will tend to yield contributions that are quadratic in the quantum fields and involve three derivatives of the appropriate quantum fields. These three-derivative contributions will destroy the minimal character of the corresponding differential operator, unless they cancel each other as in the case at hand. Notice that having two derivatives in the term in (\ref{twoderiv}) guarantees that we shall have a Laplace operator in the $\bar{c}c$ contribution to $S$ in  (\ref{quantumaction}).

It is worth stressing that the term
\be
\alpha \int d^n x \sqrt{|\bar{g}|} ~\bar{g}^{\m\n}~\partial_\m\bar{c}\,\partial_\n\bp
\ee
in  $X_W$ in (\ref{XDW}) ensures that no linear contribution in the quantum fields occurs in $s\,\left(X_D+X_W\right)$, as befits the concept of DeWitt effective action.

Let us finally point out that we shall keep the gauge parameters $\alpha$ and $\gamma$ arbitrary and thus check non-trivially that our result for the on-shell most UV contribution to $\Gamma_{DeW}[\bar{g}_{\mu\nu},\bar{\phi}]$ does not depend neither on $\alpha$ nor $\gamma$.
\par
The $B_\m$ field appears linearly and it is conveniently integrated out. It is worth remarking thet were we to do the same thing for the field $f$, the resulting operator would have been not minimal anymore.

\section{The one-loop effective action of CDG.}

With the addition of the gauge fixing term, we have all the  ingredients needed to compute the one-loop counterterm. Again, we define a generalized field living in the ''gauge" bundle, this time including also the auxiliar field $f$
\begin{align}
\Psi^{A}=\begin{pmatrix}
k^{\m\n}\\
\phi\\
f
\end{pmatrix}
\end{align}

This means that the metric $G_{AB}$ and the matrices $M_{AB}$ and $N^{\m}_{AB}$ will have now extra entries corresponding to the new interaction terms containing $f$. Therefore, the metric now reads
\begin{align}
G_{AB}=\frac{(n-2)}{4(n-1)}\begin{pmatrix}
 \frac{1}{8}{\cal G}_{\m\n\r\s}^{\a\b}\bg_{\a\b}&\frac{1}{4}\bg_{\m\n}&0 \\
\frac{1}{4} \bg_{\r\s} &\frac{n}{(n-2)}&-\frac{2\a(n-1)}{n-2}\\
0&-\frac{2\a(n-1)}{n-2}&\frac{4(n-1)}{n-2}
\end{pmatrix}
\end{align}
whose inverse, in the same sense as before, happens to be
\begin{align}
G^{AB}=\begin{pmatrix}
-\frac{16(n-1)}{n-2}\left[\bg^{\m\r}\bg^{\n\s}+\bg^{\n\r}\bg^{\m\s}+\frac{2(2+\a^{2}(1-n))}{\a^{2}(2-3n+n^{2})}\bg^{\m\n}\bg^{\r\s}   \right]&\frac{16}{\a^{2}(n-2)}\bg^{\r\s} &\frac{8}{\a (n-2)}\bg^{\r\s}\\
\frac{16}{\a^{2}(n-2)}\bg^{\m\n}&-\frac{4}{\a^{2}}&-\frac{2}{\a}\\
\frac{8}{\a (n-2)}\bg^{\m\n}&-\frac{2}{\a}&0
\end{pmatrix}
\end{align}

The matrices are extended in such a way that
\begin{align}
N^{\b}_{AB}=\begin{pmatrix}
N^{\b}_{kk}& N^{\b}_{k\phi}&  N^{\b}_{k f}\\
N^{\b}_{\f k}& N^{\b}_{\f\f} & N^{\b}_{\f f}\\
N^{\b}_{f k}&N^{\b}_{f \f}&N^{\b}_{ff}
\end{pmatrix}\quad\quad
M_{AB}=\begin{pmatrix}
M_{kk}& M_{k\phi} & M_{k f} \\
M_{\f k}& M_{\f\f} & M_{\f f}\\
M_{f k} & M_{f \f} & M_{ff}
\end{pmatrix}
\end{align}
where the $kk$, $k\f$ and $\f\f$ elements are the same as in the non-Weyl-invariant case (provided that we substitute $\xi$ by $\xi_{c}$) and the new elements read
\begin{align*}
&N^{\b}_{k f}=-N^{\b}_{f k}=\frac{\a}{4}\frac{\bD^{\a}\bf}{\bf}\left(\bg_{\a\n}\delta^{\b}_{\m}+\bg_{\a\m}\delta^{\b}_{\n}-\bg_{\m\n}\delta_{\a}^{\b}\right)\\
&N^{\b}_{ff}=0\\
&N^{\b}_{\f f}=-N^{\b}_{f\f}=0\\
&M_{k f}=M_{f k}=-\frac{\a}{8}\left(\bD_{\m}\left(\frac{\bD_{\n}\bf}{\bf}\right)+\bD_{\n}\left(\frac{\bD_{\m}\bf}{\bf}\right)-\bg_{\m\n}\bD^{\b}\left(\frac{\bD_{\b}\bf}{\bf}\right)\right)\\
&M_{\f f}=M_{f \f}=0\\
&M_{ff}=0
\end{align*}

Let us stress that now
\bea
F_{\a\b}\,^A\,_{B}&&=\begin{pmatrix}
\frac{1}{2}\left(\br^{\m}\,_{\r}\,_{\a\b}\delta^{\n}_{\s}+\br^{\n}\,_{\r}\,_{\a\b}\delta^{\m}_{\s}+\br^{\m}\,_{\s}\,_{\a\b}\delta^{\n}_{\r}+\br^{\n}\,_{\s}\,_{\a\b}\delta^{\m}_{\r}\right)&\quad\quad  0&\quad\quad 0\\
0 &\quad\quad 0&\quad\quad 0\\
0&\quad\quad 0&\quad\quad 0
\end{pmatrix}+\nonumber\\
&&\quad\quad\quad\bD_{\a}\omega_{\b}\,^{A}\,_{B}-\bD_{\b}\omega_{\a}\,^{A}\,_{B}+\omega_{\a}\,^{A}\,_{C}\omega_{\b}\,^{C}\,_{B}-\omega_{\b}\,^{A}\,_{C}\omega_{\a}\,^{C}\,_{B}
\eea

We shall introduce next the generalized ghost, $\eta^s$, and generalized antighost, $\bar{\eta}^s$, fields, which are defined as follows
\begin{align}
\eta^{s}=\begin{pmatrix} \eta^\mu\\ c \end{pmatrix}\quad\quad
\bar{\eta}^{s}=\begin{pmatrix}\bar{\eta}^\mu&\bar{c} \end{pmatrix}
\end{align}
Then, the contribution which is quadratic in the ghost and antighost fields and comes from (\ref{quantumaction}) reads
\begin{align}\label{ghostquadcon}
\hat{S}^{ghost}_2=\int d^{n}x\sqrt{|\bar{g}|}\; \bar{\eta}^{s}\left(-\bar{G}_{st}\bar{\D}^{2}+N^{\alpha}_{\phantom{\alpha}st}\bar{\D}_{\alpha}+M_{st}\right)\eta^{t},
\end{align}
where
\begin{align}\label{Gesete}
G_{st}=\begin{pmatrix}\bg_{\m\n} & 0\\0 & 1\end{pmatrix}\quad\quad
N^{\alpha}_{\phantom{\a}st}=\begin{pmatrix}
N^{\alpha}_{\phantom{\alpha}\m\n} & N^{\alpha}_{\phantom{\alpha}\m w}\\
N^{\alpha}_{\phantom{\alpha}w \n} & N^{\alpha}_{\phantom{\alpha}w w}
\end{pmatrix}\quad\quad
M_{st}=\begin{pmatrix}
M_{\m\n} & M_{\m w}\\
M_{w \n} & M_{ww}
\end{pmatrix}
\end{align}

and

\begin{align*}
&N^{\a}_{\phantom{\a}\m\n}=-(1-\gamma)\,\bar{g}_{\m\n}\,\frac{\bar{\D}^{\a}\bp}{\bp}+(1+\gamma)\,\frac{\bar{\D}_{\n}\bp}{\bp}\,\delta^{\a}_{\phantom{\a}\m}+(1-\gamma)\,\frac{\bar{\D}_{\m}\bp}{\bp}\,\delta^{\a}_{\phantom{\a}\n} \\
&N^{\a}_{\phantom{\a}\m w}=0,\quad N^{\a}_{\phantom{\a}w \nu}=\frac{2}{n-2}\frac{\bD^2\bp}{\bp}\delta^{\a}_{\phantom{\a}\nu},\quad N^{\alpha}_{\phantom{\alpha}w w}=0\\
&M_{\m\nu}=-\bar{R}_{\m\n}+2\frac{\bD_{\m}\bD_\n\bp}{\bp},\quad M_{\m w}=-\gamma (n-2) \frac{\bD_\m \bp}{\bp},\quad M_{w\n}=\frac{2}{n-2}\frac{\bD_\nu\bD^2 \bp}{\bp}\\
&M_{ww}=\frac{\bD^2\bp}{\bp}
\end{align*}
The reader should bear in mind that the index $w$ has no range and goes with $c$ or $\bar{c}$, as the case may be.

The heat kernel coefficient (\ref{heat kernel}) associated to $\hat{S}^{ghost}_2$ in (\ref{ghostquadcon}) is the corresponding coefficient of the heat kernel expansion of the following  operator
\begin{align}
\hat{\Delta}^{(ghost)} = - (\bg^{\mu\nu}[\bD_{\mu} \delta^s_{\phantom{\mu}t'}+\omega^{s}_{\mu\,t'}][\bD_{\nu} \delta^{t'}_{\phantom{\nu}t}+\omega^{t'}_{\nu\,t}]\,+\,E^{s}_{\phantom{\mu}t}),
\end{align}
where
\begin{align}
&\omega^{s}_{\mu\,t}=\frac{1}{2}G^{st'}\,N_{\mu\,t't}\\
&E^{s}_{\phantom{\mu}t}=G^{st'}(-M_{t' t}- \omega_{\m  t' s'} \omega^{\m s'}_{\phantom{\m}t}-\bD_{\m}\omega_{t' t}^{\m})
\end{align}
$G^{st}$ is the inverse matrix of $G_{st}$ in (\ref{Gesete}).

To work out the heat kernel coefficient (\ref{heat kernel}) associated to $\hat{S}^{ghost}_2$ in (\ref{ghostquadcon}), one also needs the field strength for the connection defined by $\bar{\D}_{\mu}\delta^{s}_{\phantom{s}t}+\omega_{\mu t}^{s}$, which reads
\begin{align}
&F_{\rho\sigma t}^{s}=\begin{pmatrix}
\br^{\m}_{\phantom{\m}\n\rho\sigma} & 0\\
0 & 0
\end{pmatrix}+\bD_{\rho}\omega^s_{\sigma t} - \bD_{\sigma}\omega^s_{\rho t} + [\omega_\rho,\omega_\sigma]^s_{\phantom{s}t}
\end{align}

All the ingredients  which are needed for the full computation are now in place. All that is left is to add the contributions of the physical fields and the ghost fields
\begin{align}
\Gamma_{DeW}\left[\bg,\bp\right]=\frac{1}{n-4}\left(A_{2}^{\textrm bosons}\left[\bg,\bp\right]-2~A_{2}^{\textrm ghosts}\left[\bg,\bp\right]\right)=\frac{1}{n-4}\frac{1}{16\pi^{2}}\int d^{4}x\sqrt{|\bg|}\; a_{2}\left[\bg,\bp\right]
\end{align}
where the off-shell $a_{2}$ term is now

\bea
&&a_{2}\left[\bg,\bp\right]=Q_1(\a\,\g)\frac{(\bD\bf)^{2}(\bD\bf)^{2}}{\bf^{4}}+Q_2(\a,\g)\frac{\bD^{\m}\bf\bD^{\n}\bf\bD_{\m}\bD_{\n}\bf}{\bf^{3}}+Q_3(\a,\g)\frac{\bD^{\m}\bD^{\n}\bf \bD_{\m}\bD_{\n}\bf}{\bf^{2}}+\nonumber\\
&&2\gamma\frac{\bD_{\m}\bf\bD^{2}\bD^{\m}\bf}{\bf^{2}} +Q_4(\a,\g)\frac{(\bD\bf)^{2} \bD^{2}\bf}{\bf^{3}}+Q_5(\a,\g)\frac{\bD^{2}\bf\bD^{2}\bf}{\bf^{2}}+Q_6(\a,\g)\frac{\br^{\m\n}\bD_{\m}\bf\bD_{\n}\bf}{\bf^{2}}+\nonumber\\
&&+Q_7(\a,\g)\frac{\br^{\m\n}\bD_{\m}\bD_{\n}\bf}{\bf}+Q_8(\a,\g)\br^{\m\n}\br_{\m\n}+\frac{53}{45}\br^{\m\n\a\b}\br_{\m\n\a\b}+Q_9(\a,\g)\frac{\br\bD^{\m}\bf\bD_{\m}\bf}{\bf^{2}}+\nonumber\\
&&+Q_{10}(\a,\g)\br^{2}+Q_{11}(\a,\g)\frac{\br\bD^{2}\bf}{\bf}
\eea
where
\bea
&&Q_1(\a,\g)\equiv \frac{16 + 108 \alpha^2 - 8 \gamma + 96 \alpha^2 \gamma + 4 \gamma^2 +  18 \alpha^2 \gamma^2 + \gamma^3 + 4 \alpha^2 \gamma^3}{\a^{2}}\nonumber\\
&&Q_2(\a,\g)\equiv -\frac{2 (96 + 405 \alpha^2 - 48 \gamma + 390 \alpha^2 \gamma + 13 \gamma^2 +
   57 \alpha^2 \gamma^2 + 3 \gamma^3 + 12 \alpha^2 \gamma^3)}{9 \a^{2}}\nonumber\\
&&Q_3(\a,\g)\equiv \frac{48 + 81 \alpha^2 - 24 \gamma + 102 \alpha^2 \gamma + \gamma^2 +
 3 \alpha^2 \gamma^2}{9 \a^{2}}\nonumber\\
&&Q_4(\a,\g)\equiv -\frac{-102 - 378 \alpha^2 + 96 \gamma - 420 \alpha^2 \gamma - 44 \gamma^2 -
 60 \alpha^2 \gamma^2 + 3 \gamma^3 + 12 \alpha^2 \gamma^3}{9\a^{2}}\nonumber\\
&&Q_5(\a,\g)\equiv -\frac{-162 + 228 \alpha^2 - 108 \alpha^4 - 24 \alpha^2 \gamma +
 84 \alpha^4 \gamma + \alpha^2 \gamma^2 + 3 \alpha^4 \gamma^2}{9 \a^{4}}\nonumber\\
&&Q_6(\a,\g)\equiv -\frac{-96 - 63 \alpha^2 + 24 \gamma - 78 \alpha^2 \gamma - \gamma^2 +
 15 \alpha^2 \gamma^2}{9 \a^{2}}\nonumber\\
&&Q_7(\a,\g)\equiv -\frac{4 (4 - 3 \alpha^2 - \gamma + \alpha^2 \gamma)}{3 \a^{2}}\nonumber\\
&& Q_8(\a,\g)\equiv -\frac{-120 + 361 \alpha^2}{90 \a^{2}}\nonumber\\
&&Q_9(\a,\g)\equiv -\frac{11 + 24 \alpha^2 - 6 \gamma + 32 \alpha^2 \gamma + 3 \gamma^2 +
6 \alpha^2 \gamma^2}{3 \a^{2}}\nonumber\\
&&Q_{10}(\a,\g)\equiv \frac{18 - 30 \alpha^2 + 43 \alpha^4}{36 \a^{4}}\nonumber\\
&&Q_{11}(\a,\g)\equiv \frac{-18 + 25 \alpha^2 - 21 \alpha^4 - 2 \alpha^2 \gamma + 2 \alpha^4 \gamma}{3\a^{4}}\nonumber
 \eea
 It is worth mentioning that all the monomials including the scalar field diverge when $\bp=0$. Naive power counting arguments can not then be applied. This fact also prevents the monomials that appear in the bare lagrangian to appear in the counterterm. This physically means something that we already knew, namely that our calculation is restricted to the broken phase of the theory.
When this is put on-shell by using the relationships derived in a previous paragraph (particularized for the conformal value of the coupling $\xi$) all the gauge dependence on the parameters $\gamma$ and $\alpha$ dissapears. This is a powerful check of the gauge independence of our result. Moreover, by using the relations
\begin{align}
&E_{4}=\br_{\m\n\a\b}\br^{\m\n\a\b}+\br^{2}-4\br_{\m\n}\br^{\m\n}\\
&\int ~d(vol)\;E_{4}=\int ~d(vol)\;W_{4}-2\int ~d(vol) \;\left(\br^{\m\n}\br_{\m\n}-\frac{1}{3}\br^{2}\right)
\end{align}
and using the fact that the last term in the second relation vanishes when using the EM, the on-shell counterterm  finally reads
\begin{align}
\Gamma_{DeW}=\frac{1}{n-4}\frac{1}{16\pi^{2}}\frac{53}{45}\int ~d(vol)\;E_{4}
\end{align}
which is exactly the same as in General Relativity once the equations of motion are taken into account. In fact the counterterm vanishes for manifolds with vanishing Euler characteristic (although not in general). On the other hand, S-matrix elements depend only on the on-shell effective action. This means that the anomaly induced by the corresponding evanescent operator is trivial for those spaces with $\chi(M)=0$.
\section{Inclusion of a quartic interaction.}
The attentive reader could not fail to notice that the action of CDG is not  the most general one with the full set of symmetries. There is an operator, namely a quartic self-interaction of the graviscalar field, that can always be included. The reason why it has been taken apart from the other terms will be apparent in a moment.
\par
The action of interest is 
\begin{align}
S=-\int d^{n}x ~\sqrt{|g|}~\left(\xi \phi^2 R + \frac{1}{2}\D_{\m}\f\D^{\m}\f - \lambda \f^{\frac{2n}{n-2}}\right)
\end{align}
where  the coupling constant is dimensionless in any dimension.

The presence of this new term will add corrections to the second order action when expanding around background fields. All these terms will be just new additions to the $M_{AB}$ matrix of our algorithm
\begin{align}
M_{AB}^{\lambda}=M^{0}_{AB}+\delta M_{AB}
\end{align}
where $M_{AB}^{0}$ is the corresponding matrix when $\lambda$ vanishes and
\begin{align}
\delta M_{AB}=\begin{pmatrix}
\frac{\lambda}{4}\bar{\f}^{\frac{4}{n-2}}\left({\cal P}_{\m\n\r\s}^{\a\b}-\frac{1}{2}{\cal K}_{\m\n\r\s}^{\a\b}\right) \quad& -\frac{\lambda}{2}\frac{n}{n-2} \bar{\f}^{\frac{4}{n-2}}\\ 
-\frac{\lambda}{2}\frac{n}{n-2} \bar{\f}^{\frac{4}{n-2}} & -\lambda\frac{n(n+2)}{(n-2)^{2}}\bar{\f}^{\frac{4}{n-2}} 
\end{pmatrix}
\end{align}
It is convenient to study separately the non-conformal and the conformal case.
\subsection{Non-conformal Dilaton Gravity}
Let us start with Dilaton Gravity out of the conformal point and just repeat the steps we did before to obtain the one-loop effective action. For simplicity we choose $\gamma=0$ and obtain the following off-shell result after performing the computations for the DeWitt-Schwinger coefficient of the Heat Kernel (the corresponding ghost action is insensible to the adition of the potential term)
\begin{align}
a_{2}^{\lambda}=a_{2}^{0}+\delta a_{2}
\end{align}
where, again, $a_{2}^{0}$ is the coefficient when the potential is absent, obtained by choosing the $\gamma =0$ gauge in the corresponding equation, and
\be
\delta a_{2}=l_1(\g,\l,\xi) (\D\bf)^{2}+l_2(\g,\l,\xi)\phi^{4} -l_3(\g,\l,\xi) \bar{R}\bf^{2}- l_4(\g,\l,\xi)\bf\D^{2}\bf 
\ee
All counterterms proportional to the quartic self-interaction renormalize coupling constants already present in the original lagrangian; in particular there are no operators that become singular when $\bp\rightarrow 0$
\par
Here
\bea
&&l_1(\g,\l,\xi)=2880 (-1 + 12 \Ep) (1 - \Ep (25 + 18 \gamma + 5 \gamma^2) + 
   4 \Ep^2 (45 + 50 \gamma + 18 \gamma^2))\lambda\nonumber\\
&&l_2(\g,\l,\xi)=2880 (5 - 112 \Ep + 696 \Ep^2) \lambda^2 \nonumber\\
&&l_3(\g,\l,\xi)=960 \Ep (13 - 282 \Ep + 1728 \Ep^2)\lambda\nonumber\\
&&l_4(\g,\l,\xi)=5760 \Ep (-7 + \Ep (132 - 46 \gamma) + 264 \Ep^2 (-3 + \gamma) + 2 \gamma)\lambda
\eea

The presence of the self-interaction also corrects the equations of motion for the background fields. They receive new pieces and now read in four dimensions
\begin{align*}
&\xi R_{\m\n}=\frac{1}{4}g_{\m\n}\frac{\D^{2}\bf}{\bf}-\left(\frac{1}{2}-2\xi\right)\frac{\D_{\m}\bf\D_{\n}\bf}{\bf^{2}}-\left(2\xi-\frac{1}{4}\right)g_{\m\n}\frac{(\D\bf)^{2}}{\bf^{2}}+2\xi\frac{\D_{\m}\D_{\n}\bf}{\bf}-2\xi g_{\m\n}\frac{\D^{2}\bf}{\bf}+\frac{1}{2}\lambda \bf^{2}g_{\m\n}\\
&R-\frac{1}{2\xi} \frac{\D^{2}\bf}{\bf}-\frac{2\lambda}{\xi}\bf^{2}=0
\end{align*}

 The second set of on-shell relations previously derived in the absence of self-interaction are still valid, since they only involve integration by parts. The first set is however modified by the presence of the self-interaction. They read now
\begin{align*}
&\frac{\D^{2}\bf\bar{R}}{\bf}=\frac{1}{2\xi}A-\frac{2\lambda}{\xi}(\D\bf)^{2}\\
&\frac{(\D\bf)^{2}\bar{R}}{\bf}=\frac{1}{2\xi}B+\frac{2\lambda}{\xi}(\D\bf)^{2}\\
&\bar{R}^{2}=\frac{1}{4\xi^{2}}A+\frac{4\lambda^{2}}{\xi^{2}}\bf^{4}+\frac{2\lambda}{\xi^{2}}\bf\D^{2}\bf\\
&\frac{\bar{R}^{\m\n}\D_{\m}\bf\D_{\n}\bf}{\bf^{2}}=\left(\frac{1}{4\xi}-2\right)B-\frac{1}{4\xi}C+2E+\frac{\lambda}{2\xi}(\D\bf)^{2}\\
&\frac{\bar{R}^{\m\n}\D_{\m}\D_{\n}\bf}{\bf}=\left(\frac{1}{4\xi}-2\right)(A+B)+2D+\left(2-\frac{1}{2\xi}\right)E+\frac{\lambda}{2\xi}\bf\D^{2}\bf\\
&\bar{R}^{\m\n}\bar{R}_{\m\n}=\left(\frac{1}{4\xi}-2\right)\left(\bar{R}\frac{\D^{2}\bf}{\bf}+\bar{R}\frac{(\D\bf)^{2}}{\bf^{2}}\right)+\left(2-\frac{1}{2\xi}\right)\frac{\bar{R}^{\m\n}\D_{\m}\bf\D_{\n}\bf}{\bf^{2}}+2\frac{\bar{R}^{\m\n}\D_{\m}\D_{\n}\bf}{\bf}+\frac{\lambda}{2}\bar{R}\bf^{2}
\end{align*}  
And, again as in the $\lambda =0$ case, compatibility of the two equations, whenever we are out of the conformal point, demands
\begin{align}
A=C=-B
\end{align}

Using this modified relations, we can finally put the full counterterm on-shell, ending up with a simple expression for the DeWitt-Schwinger  coefficient
\begin{align}
a_{2}^{\lambda(on-shell)}=\frac{71}{60}W_{4}+\frac{1259}{1440}(1-12\xi)^{2}C+\frac{1484}{1440}\frac{1-12\xi}{\xi^{2}}\lambda(\D\bf)^{2}-\frac{371}{180}\frac{\lambda^{2}}{\xi^{2}}\bf^{4}
\end{align}

The main physical effect of the self-interaction  at this level is to  generate counterterms  for the dimension four operators  in the lagrangian, a feature that was absent before. Actually   the renormalization of  the non-minimal coupling to the curvature is proportional to
\be
\delta \xi \propto \frac{\lambda}{\xi^{2}}.
\ee
In the limit $\xi \rightarrow \frac{1}{12}$, corresponding to the conformal value, all non-Weyl invariant terms in the effective action vanish. However, as we already saw in the pure $\lambda=0$ case, this limit is discontinuous owing to the presence of an enhanced gauge symmetry so that the coefficients in front of every term will be different  in the conformal case.
\subsection{Conformal Dilaton Gravity}
Let us now turn our attention to the conformal  case in which the coupling to curvature reaches the conformal value $\xi_{c}=\frac{1}{12}$. In this case, we have a Conformal Dilaton Gravity with an extra gauge symmetry, namely Weyl invariance. The only monomial compatible with this new symmetry is precisely 
\begin{align}
V=\lambda \f^{\frac{2n}{n-2}}
\end{align}
Actually it is the only  Weyl invariant potential term in arbitrary dimension.
\par
Quantization of CDG in this phase is done, again, in the same way as in the $\lambda=0$ case. We stick to the $\gamma=0$ choice for the diffeomorphism gauge fixing and we introduce the gauge fixing sector for Weyl invariance by using BRS techniques as before. In this case, the matrix $\delta M_{AB}$ must be extended to include the $f$ field in a trivial way as
\begin{align}
\delta M_{AB}=\begin{pmatrix}
\frac{\lambda}{4}\bar{\f}^{\frac{4}{n-2}}\left({\cal P}_{\m\n\r\s}^{\a\b}-\frac{1}{2}{\cal K}_{\m\n\r\s}^{\a\b}\right) \quad& -\frac{\lambda}{2}\frac{n}{n-2} \bar{\f}^{\frac{4}{n-2}} \quad&0 \\ 
-\frac{\lambda}{2}\frac{n}{n-2} \bar{\f}^{\frac{4}{n-2}} & -\lambda\frac{n(n+2)}{(n-2)^{2}}\bar{\f}^{\frac{4}{n-2}} \quad &0 \\ 
 0&  0& 0
\end{pmatrix}
\end{align}

Plugging the new matrix into the algorithm and working out the computations, the off-shell Heat kernel expression again receives new terms proportional to $\lambda$. These are, when $n=4$
\be
\delta a^{\lambda}_{2}=s_1(\a,\l)~ \bf^{4}+ s_2(\a,\l) (\D\phi)^2+ s_3(\a,\l) \bp\D^{2}\bp- s_4(\a,\l) \bar{R}\bf^{2}
\ee
It is remarkable that also here all counterterms involving $\l$ seem to obey the ordinary power counting arguments and renormalize the coupling constants already present in the bare lagrangian. No singularities when $\bp\rightarrow 0$ are present in terms involving the quartic self-interaction.
\par
The explicit values of the four functions $s_i(\a,\l)$ is given by
\bea
&&s_1(\a,\l)\equiv 48\lambda^2 \frac{6 - 4 \alpha^2 + 15 \alpha^4}{\a^4}\nonumber\\
&&s_2(\a,\l)\equiv 12\lambda \left(3 + \frac{2}{\alpha^2}\right)\nonumber\\
&&s_3(\a,\l)\equiv 24 \lambda\frac{6 - 6 \alpha^2 + 7 \alpha^4}{ \alpha^4}\nonumber\\
&&s_4(\a,\l)\equiv 4\lambda \frac{6 - 5 \alpha^2 + 13 \alpha^4}{ \alpha^4}
\eea
So that using again the on-shell relations induced by the equations of motion, we find the on-shell coefficient to be
\begin{align}
a_{2}^{\lambda(on-shell)}=\frac{53}{45}W_{4}-\frac{4568}{15}\lambda^{2}\bf^{4}
\end{align}

\par
A quartic self-interaction in the Jordan frame corresponds to a cosmological constant in the Einstein frame. It must be then the case that the counterterm just derived is the Weyl transformation of the one obtained for General Relativity with a cosmological constant by Christensen and Duff \cite{Christensen}. 
\par
 This counterterm reads on-shell
\begin{align}
a_{2}^{GR(on-shell)}=\frac{53}{45}W_{4}-\frac{1142}{135}\Lambda^{2}
\end{align}

If the above conjecture is to be true our $\lambda$ must be directly related to their $\Lambda$.

Taking the limit in which CDG goes to General Relativity, characterized in n=4 dimensions by 
\be
\f\rightarrow\sqrt{12}\;M_{p}
\ee
we learn that 
 \be
 \Lambda = 6 \lambda \f^{2}
 \ee
The Christensen-Duff counterterm then reads
\begin{align}
a_{2}^{GR(on-shell)}\longrightarrow \frac{53}{45}W_{4}-\frac{4568}{15}\lambda^{2}\bf^{4}
\end{align}
which is exactly the result we obtained by a direct computation of the one-loop counterterm of CDG.

\par
This is perhaps a good place to comment somewhat on previous literature. In \cite{BarvinskyKK} some similar models are analyzed; but they beg the main physical question in the sense that they postulate that the counterterm should be the Weyl transform of the 't Hooft and Veltman's one.
\par
The are a couple of interesting papers (\cite{Shapiro}\cite{Steinwachs}) where quite general models  that include the one studied in the present paper are analyzed outside the conformal point.  Only \cite{Shapiro} reports on shell results, so that there can be a meaningful comparison. We have checked that the coefficent of $W_4$ in their on-shell counterterm is different from our result.
\par
The other paper \cite{Steinwachs} also assumes from the start that the function $U(\phi)$ that multiplies the scalar curvature cannot vanish\footnote{We are grateful to A. Kamenshchik for informing us of this fact.}. This means that their results do not hold when $U(\phi)\sim \phi^2$ as in our case. No comparison can then be made with them.
\section{Physical effects of quantum  gravity.}
The standard lore of effective field theories is that quantum gravity effects should decouple at energies much smaller than Planck mass, $M_p\equiv {1\over \sqrt{16 \pi G}}$, so that they can be safely ignored in particle physics except in exotic circumstances.
This statement needs qualification in all cases in which the gravitational coupling constant becomes dynamical. This is what happens, in particular, in conformally invariant theories, where all energy scales are physically equivalent. Our calculations as reported here allow for a quantitative axample.
\par
An scalar field conformally coupled to the gravitational field has the action
\be
S=\int d(vol)\left({n-2\over 8(n-1)}\phi^2 R+{1\over 2}\left(\nabla\phi\right)^2\right)
\ee
The conformal invariance of the effective action implies an off-shell  Ward identity  
\be\label{cwi}
2 g^{\m\n}{\d \Gamma\over \d g^{\m\n}}+{n-2\over 2}\phi{\d \Gamma\over \d\phi}=0
\ee
This is true irrespectively of whether gravitation is dynamical or not.
\par
The corresponding energy momentum tensor of the scalar fields reads
\bea
&&T_{\a\b}\equiv {2\over \sqrt{|g|}}{\d S\over \d g^{\a\b}}={n-2\over 4(n-1)}~R_{\a\b}\phi^2+{n\over 2(n-1)}\nabla_\a\phi\nabla_\b\phi-{n-2\over 2(n-1)}\phi\nabla_\a\nabla_\b\phi-\nonumber\\
&&-{1\over 2}\left({n-2\over 4(n-1)} R \phi^2+{1\over n-1} (\nabla\phi)^2-{n-2\over n-1}\phi\nabla^2\phi\right)~g_{\a\b}
\eea
This energy-momentum tensor is already  traceless on shell
\be
T\equiv g^{\a\b}T_{\a\b}=0.
\ee
Quantum corrections yield however a trace anomaly ({\em confer}, for example \cite{Parker}, page 107) which in this case is given by

\be
\left\langle 0\left|T\right|0\right\rangle={1\over 2880\pi^2} \left({3\over 2}~W_4 - {1\over 2}~ \bar{E}_4 + \Box~\bar{R}\right)
\ee
Not only that; even the Ward identity [\ref{cwi}] is violated as well owing to evanescent operators, acquiring a nonvanishing second member
\bea\label{windg}
&&2 g^{\m\n}{\d \Gamma\over \d g^{\m\n}}+\phi{\d \Gamma\over \d\phi}=\frac{1}{2880\pi^{2}}\left(\frac{3}{2}\bar{W}_{4}-\frac{1}{2}\bar{E}_{4}+\bar{\Box} \bar{R}\right)+\nonumber\\
&&+\frac{9}{8\pi^{2}}\lambda^{2}\bp^{4}-\frac{\lambda}{8\pi^{2}} \D_{\m}\left(\bp\D^{\m}\bp\right)
\eea
The quartic self-interaction for the scalar field has been included.
\par
This is to be contrasted with the result just obtained  when gravitation is dynamical
\be
2 g^{\m\n}{\d \Gamma\over \d g^{\m\n}}+\phi{\d \Gamma\over \d\phi}={1\over 16\pi^2}\left(\frac{53}{45}\bar{W}_{4}-\frac{4568}{15}\lambda^{2}\bf^{4}\right)
\ee
Here $\Gamma$ is the four-dimensional renormalized action, and the quartic self-interaction has been included as well.
 The difference between the result in the presence of quantum gravity effects and the result when the gravitational field is just a background is not small. This is only natural, because there is no yet anything that fixes the scale at which quantum gravity effects should become important.

\newpage
\section{A discussion of the fate of the Weyl symmetry Ward identity at the two loop level}
The fact that the UV divergent counterterm (as well as the conformal anomaly) vanishes on shell  means that it is in principle irrelevant, at least as far as S-matrix physics is concerned. It is most interesting to consider a situation in which UV divergences are likely to show up. As we shall see, this is the case in four dimensions at the two loop order and six dimensions at the one loop order. Let us begin with the case in four dimensions.

The fact that the action $S_{CDG}$ in (\ref{Scdg}) is Weyl invariant for arbitrary $n$ and the results concerning the Quantum Action Principle presented in \cite{Breitenlohner:1977hr} lead to the conclusion that the dimensionally regularized {\em on shell} background field effective action,
$\Gamma[\bar{g}_{\mu\nu},\bp;\,n]$,  of our theory is Weyl invariant at any loop order:
\be\label{regwardid}
\Big[2\,\bar{g}^{\mu\nu}\,{\delta\phantom{\bar{g}^{\mu\nu}}\over\delta\bar{g}^{\mu\nu}}+{n- 2 \over 2}\bp\,{\delta\phantom{\bp}\over\delta\bp}\,\Big]\Gamma[\bar{g}_{\mu\nu},\bp;\,n]\,=\,0.
\ee

And yet --see Theorem 2 of \cite{Breitenlohner:1977hr}, that the previous equation holds does not necessarily mean that the renormalized effective action, say $\Gamma_0[\bar{g}_{\mu\nu},\bp]$, obtained from $\Gamma[\bar{g}_{\mu\nu},\bp;\,n]$ by using the minimal substraction algorithm satisfies the corresponding Ward identity in four dimensions. Indeed, let us assume --an assumption to be discussed below-- that $\Gamma[\bar{g}_{\mu\nu},\bp;\,n]$ develops a simple pole at two loops; then, since the Ward identity in (\ref{regwardid}) contains coefficients with an explicit dependence on $n$, these coefficients  may give rise to contributions that cancel the pole at $n-4$ in $\Gamma[\bar{g}_{\mu\nu},\bp;\,n]$. This mechanism may yield UV finite terms --let us denote them by
 ${\cal B}[\bar{g}_{\mu\nu},\bp]$-- that break the Ward identity for $\Gamma_0[\bar{g}_{\mu\nu},\bp]$:
 \be\label{anoma}
\Big[2\,\bar{g}^{\mu\nu}\,{\delta\phantom{\bar{g}^{\mu\nu}}\over\delta\bar{g}^{\mu\nu}}+\bp\,{\delta\phantom{\bp}\over\delta\bp}\,\Big]
\Gamma_0[\bar{g}_{\mu\nu},\bp]\,=\,{\cal B}[\bar{g}_{\mu\nu},\bp].
\ee
 The previous Ward identity breaking term, ${\cal B}[\bar{g}_{\mu\nu},\bp]$,  will turn to be a true anomaly if no acceptable UV finite counterterm, $\Gamma_0^{(ct)}[\bar{g}_{\mu\nu},\bp]$, can be found so that
\be
\Big[2\,\bar{g}^{\mu\nu}\,{\delta\phantom{\bar{g}^{\mu\nu}}\over\delta\bar{g}^{\mu\nu}}+\bp\,{\delta\phantom{\bp}\over\delta\bp}\,\Big]
\Gamma_0^{(ct)}[\bar{g}_{\mu\nu},\bp]\,=\,{\cal B}[\bar{g}_{\mu\nu},\bp]
\ee
holds.

Let us stress that to tell whether or not ${\cal B}[\bar{g}_{\mu\nu},\bp]$ is a true anomaly, one should define first what an acceptable counterterm is. In theories with overall UV divergences which are polynomials
in the fields and their derivatives, by acceptable counterterms one means polynomials in the fields and their derivatives of the appropriate mass dimension. The reader should also bear in mind that if a true anomaly
does not show up after performing a minimal substraction, then, a true anomaly cannot be generated by performing any acceptable non-miminal substraction. By an acceptable  non-minimal substraction one means that which  differs from the minimal subtraction by acceptable UV finite  counterterms. Notice, however, that the  value of ${\cal B}[\bar{g}_{\mu\nu},\bp]$ changes, in general, as we change the acceptable substraction.

Now, in keeping with the one loop result obtained above, we shall assume that at two loops the pole part at $n=4$ of our dimensionally regularized  {\em on shell} background field effective action,  $\Gamma[\bar{g}_{\mu\nu},\bp;\,n]$, can be obtained, by performing an appropriate Weyl transformation, from the {\em on shell} two loop result worked out  in 1986 for General Relativity by Goroff and Sagnotti \cite{Goroff}. This {\em on shell} two loop divergence reads
\be\label{gorsag}
\Gamma_\infty^{(GS)}[\bar{G}_{\mu\nu}]={1\over n-4}{1\over \left(4\pi\right)^4 M_p^2}{209\over 2880} \int d^4 x \sqrt{|\bar{G}|}\,W_6^{(4)}[\bar{G}_{\mu\nu}],
\ee
where
\be
W_6^{(4)}\equiv W^{(4)\,\a_1\a_2\a_3\a_4}~ W^{(4)}_{a_3\a_4\a_5\a_6}~ W^{(4)\,a_5\a_6}\,_{\a_1\a_2}.
\ee
The symbol $W^{(4)}_{\mu_1\mu_2\mu_3\mu_4}$ stands for the Weyl tensor -see (\ref{weylten})-- for the metric $\bar{G}_{\mu\nu}$ for $n=4$.

Now, by applying the Weyl transformation
\be\label{4weyltrans}
\bar{G}_{\mu\nu}=\frac{1}{12\,M_p^2}\,\bp^2\,\bar{g}_{\mu\nu}
\ee
to $\Gamma_\infty^{(GS)}[\bar{G}_{\mu\nu}]$ in (\ref{gorsag}), one obtains
\be\label{UVdiv}
\Gamma_\infty^{(GS)}[\bar{g}_{\mu\nu},\bp]={1\over n-4}{12\over \left(4\pi\right)^4 }{209\over 2880} \int d^4 x \sqrt{|\bar{g}|}\, \frac{1}{\bp^2}\,W_6^{(4)}[\bar{g}_{\mu\nu}],
\ee
which, as stated above, we assume it is the two loop pole part contribution to the dimensionally regularized {\em on shell} background field effective action of our theory.

Since we want to make sure that no diffeomorphism anomaly arises in the renormalized theory in 4 dimensions, the
substraction of $\Gamma_\infty^{(GS)}[\bar{g}_{\mu\nu},\bp]$  in (\ref{UVdiv}) from $\Gamma[\bar{g}_{\mu\nu},\bp;\,n]$ is to be done in such a  way  that it preserves explicitly invariance under diffeomorphisms in $n$ dimensions. This is achieved by generalizing $\Gamma_\infty^{(GS)}[\bar{G}_{\mu\nu}]$ from 4 dimensions to $n$ dimensions and subtracting the resulting term from the dimensionally regularized action. The geometrically natural generalization of $W_6^{(4)}$ to the $n$ dimensional space is obtained by using the both the metric and the Weyl tensor in $n$ dimensions. The Weil tensor in $n$ dimensions contains coefficients that depend explicitly on $n$; so this generalization of $W^{(4)}_{\mu_1\mu_2\mu_3\mu_4}$ to an object in $n$ dimensions will lead to a non-minimal substraction algorithm. However, it is the Weyl tensor in $n$ dimensions the object which supplies a Weyl invariant tensor in $n$ dimensions: a property much appreciated if one looks for Weyl invariance.

Then, let us introduce the following generalization of $\Gamma_\infty^{(GS)}[\bar{g}_{\mu\nu},\bp]$  in (\ref{UVdiv}) to $n$ dimensions:
\be
\Gamma_\infty^{(nm)}[\bar{g}_{\mu\nu},\bp]=\frac{1}{n-4}\,W_{-1}^{(nm)}[\bar{g}_{\mu\nu},\bp],
\ee
where
\be
W_{-1}^{(nm)}[\bar{g}_{\mu\nu},\bp]={12\over \left(4\pi\right)^4 }{209\over 2880} \int d^n x \sqrt{|\bar{g}|}\,    {1\over \bp^2}~W_6^{(n)}[\bar{g}_{\mu\nu}].
\ee
In the previous equation
\be\label{wsixn}
W_6^{(n)}[\bar{g}_{\mu\nu}]\equiv W^{\a_1\a_2\a_3\a_4}~ W_{\a_3\a_4\a_5\a_6}~ W^{\a_5\a_6}\,_{\a_1\a_2},
\ee
$W_{\mu_1\mu_2\mu_3\mu_4}$ being  the Weyl tensor for the metric $\bar{g}_{\mu\nu}$ in $n$ dimensions --see definition in (\ref{weylten}).

We define, up to two loops, a  renormalized {\em on shell} background field effective action, let us call it $\Gamma_0^{(nm)}[\bar{g}_{\mu\nu},\bp]$, by performing the following non-minimal subtraction:
\be\label{nmaction}
\Gamma_{0}^{(nm)}[\bar{g}_{\mu\nu},\bp]=\lim_{n\rightarrow 4}\{\,\Gamma[\bar{g}_{\mu\nu},\bp;\,n]- \frac{1}{n-4}\,W_{-1}^{(nm)}[\bar{g}_{\mu\nu},\bp]\}.
\ee

Substituting
\be
\Gamma[\bar{g}_{\mu\nu},\bp;\,n]=\frac{1}{n-4}\,W_{-1}^{(nm)}[\bar{g}_{\mu\nu},\bp]\,+\,\Gamma_{0}^{(nm)}[\bar{g}_{\mu\nu},\bp]\,+\,O(n-4)
\ee
in (\ref{regwardid}) and taking into account that
\be
\Big[2\,\bar{g}^{\mu\nu}\,{\delta\phantom{\bar{g}^{\mu\nu}}\over\delta\bar{g}^{\mu\nu}}+{n- 2 \over 2}\bp\,{\delta\phantom{\bp}\over\delta\bp}\,\Big]W_{-1}^{(nm)}[\bar{g}_{\mu\nu},\bp]=
-2(n-4)\,W_{-1}^{(nm)}[\bar{g}_{\mu\nu},\bp],
\ee
one readily shows that  $\Gamma_0^{(nm)}[\bar{g}_{\mu\nu},\bp]$ satisfies the following broken Ward identity
\be\label{nmbreak}
\Big[2\,\bar{g}^{\mu\nu}\,{\delta\phantom{\bar{g}^{\mu\nu}}\over\delta\bar{g}^{\mu\nu}}+\bp\,{\delta\phantom{\bp}\over\delta\bp}\,\Big]
\Gamma_0^{(nm)}[\bar{g}_{\mu\nu},\bp]\,=\,{\cal B}[\bar{g}_{\mu\nu},\bp]=2\,W_{-1}[\bar{g}_{\mu\nu},\bp].
\ee
$W_{-1}[\bar{g}_{\m\n},\bp]$ is given by
\be\label{Wminusone}
W_{-1}[\bar{g}_{\mu\nu},\bp]={12\over \left(4\pi\right)^4 }{209\over 2880} \int d^4 x \sqrt{|\bar{g}|}\,    {1\over \bp^2}~W_6^{(4)}[\bar{g}_{\mu\nu}].
\ee

It is clear that $W_{-1}[\bar{g}_{\mu\nu},\bp]$ cannot be canceled by adding to $\Gamma_0^{(nm)}[\bar{g}_{\mu\nu},\bp]$ an integrated local polynomial of the fields and their derivatives. This is not surprising since after all the pole part of $\Gamma[\bar{g}_{\mu\nu},\bp;\,n]$ is not a polynomial  in $\bp$; although it is a meromorphic function, when $\bp$ is replaced by a complex variable. To  gain some understanding on the
type of counterterms that one has to accept with the purpose of modifying the value of Ward identity breaking term ${\cal B}[\bar{g}_{\mu\nu},\bp]$,
and eventually setting it to zero, we shall consider the effect on ${\cal B}[\bar{g}_{\mu\nu},\bp]$ of another non-minimal subtraction. This
subtraction has the same dependence on  the Weyl tensor and the metric as the previous one, but a involves non-homomorphic function of $\bp$.

Following the ideas presented in \cite{Englert}, we shall introduce first the  following non-minimal generalization to $n$ dimensions of the Goroff and Sagnotti UV divergence in (\ref{gorsag}):
\be
\Gamma_\infty^{(GS)}[\bar{G}_{\mu\nu};\,n]={1\over n-4}{1\over \left(4\pi\right)^4 M_p^{(6-n)}}{209\over 2880} \int d^n x \sqrt{|\bar{G}|}\,W_6[\bar{G}_{\mu\nu}].
\ee
Notice that $\bar{G}_{\mu\nu}$ and the Weyl tensor in $W_6[\bar{G}_{\mu\nu}]$ live in $n$ dimensions. Then, the following Weyl transformation in $n$ dimensions
\be\label{nweyltrans}
\bar{G}_{\m\n}\equiv ~{1\over M_p^2}\left({n-2\over 8(n-1)}\right)^{2\over n-2}~\big(\bp\big)^{4\over n-2}~\bar{g}_{\m\n}
\ee
casts $\Gamma_\infty^{(GS)}[\bar{G}_{\mu\nu};\,n]$ into the form
\be
\Gamma_\infty^{(wi)}[\bar{g}_{\mu\nu},\bp;\,n]=\frac{1}{n-4}{1\over \left(4\pi\right)^4 }{209\over 2880}\,\Big(\frac{n-2}{8(n-1)}\Big)^{\frac{n-6}{n-2}}\,
 \int d^n x \sqrt{|\bar{g}|}\,   \big(\bp\big)^{2\frac{n-6}{n-2}}~W_6^{(n)}[\bar{g}_{\m\n}].
\ee

The renormalized, $\Gamma_0^{(wi)}[\bar{g}_{\mu\nu},\bp]$, {\em on shell} background field effective action, up two loops, is now defined by using
$\Gamma_\infty^{(wi)}[\bar{g}_{\mu\nu},\bp;\,n]$ to implement the following non-minimal substraction
\be\label{wiaction}
\Gamma_{0}^{(wi)}[\bar{g}_{\mu\nu},\bp]=\lim_{n\rightarrow 4}\{\,\Gamma[\bar{g}_{\mu\nu},\bp;\,n]-
\Gamma_\infty^{(wi)}[\bar{g}_{\mu\nu},\bp;\,n]\}.
\ee
Finally, the fact that $\Gamma_\infty^{(wi)}[\bar{g}_{\mu\nu},\bp;\,n]$ is Weyl invariant in $n$ dimensions --it satisfies (\ref{regwardid})--
and that now
\be
\Gamma[\bar{g}_{\mu\nu},\bp;\,n]=\Gamma_\infty^{(wi)}[\bar{G}_{\mu\nu};\,n]\,+\,\Gamma_{0}^{(wi)}[\bar{g}_{\mu\nu},\bp]\,+\,O(n-4)
\ee
leads to the conclusion that $\Gamma_0^{(wi)}[\bar{g}_{\mu\nu};\bp]$ is Weyl invariant:
\be
\Big[2\,\bar{g}^{\mu\nu}\,{\delta\phantom{\bar{g}^{\mu\nu}}\over\delta\bar{g}^{\mu\nu}}+\bp\,{\delta\phantom{\bp}\over\delta\bp}\,\Big]
\Gamma_0^{(wi)}[\bar{g}_{\mu\nu},\bp]\,=\,0.
\ee

We would like to point out that to obtain an {\em on shell} renormalized effective action that is Weyl invariant, we have substracted an integrated
function which contains, for complex $n$ close to 4, a non-meromorphic function of $\bp$; namely, $\big(\bp\big)^{2\frac{n-6}{n-2}}$. Non-surprisingly, the Laurent expansion of $\Gamma_\infty^{GS}[\bar{g}_{\mu\nu},\bp;\,n]$ around $n=4$ contains the non-zero UV finite --ie, non-vanishing in the limit $n\rightarrow 4$-- term
\be\label{nhterm}
W_0^{(nh)}[\bar{g}_{\m\n},\bp]={12\over \left(4\pi\right)^4 }{209\over 2880} \int d^4 x \sqrt{|\bar{g}|}\,    {1\over \bp^2}~
\ln(\bp)^2~W_6^{(4)}[\bar{g}_{\mu\nu}],
\ee
which, in turns, contains the non-meromorphic --when $\bp$ is replaced by a complex variable-- logarithm. Whether substractions
 involving such terms are acceptable to define a quantum field theory of gravity is an open issue, which we will not discuss in this paper.

 Notice that
\be\label{nhbreak}
\Big[2\,\bar{g}^{\mu\nu}\,{\delta\phantom{\bar{g}^{\mu\nu}}\over\delta\bar{g}^{\mu\nu}}+\bp\,{\delta\phantom{\bp}\over\delta\bp}\,\Big]
W_0^{(nh)}[\bar{g}_{\mu\nu},\bp]\,=\,2\,W_{-1}[\bar{g}_{\m\n},\bp],
\ee
where $W_{-1}[\bar{g}_{\m\n},\bp]$ is given in (\ref{Wminusone}). Hence,  by finite renormalizing $\Gamma_{0}^{(nm)}[\bar{g}_{\mu\nu},\bp]$ in
(\ref{nmaction}) as follows
\be
\Gamma_{0}^{(nm)}[\bar{g}_{\mu\nu},\bp]\rightarrow \Gamma_{0}^{(new)}[\bar{g}_{\mu\nu},\bp]=
\Gamma_{0}^{(nm)}[\bar{g}_{\mu\nu},\bp]\,-\,W_0^{(nh)}[\bar{g}_{\m\n},\bp],
\ee
one obtains a renormalized {\em on shell} background field effective action which is Weyl invariant. Of course, all this is a consequence of the fact that the difference between  the substraction term, $\Gamma_\infty^{(wi)}[\bar{g}_{\mu\nu},\bp;\,n]$, used in (\ref{wiaction}) and the substraction term, $1/(n-4)\,W_{-1}^{(nm)}[\bar{g}_{\mu\nu},\bp]$, employed in (\ref{nmaction}) contains $W_0^{(nh)}[\bar{g}_{\mu\nu},\bp]$ in (\ref{nhterm}), in the limit $n\rightarrow 4$.

The outcome of the analysis and computations we have carried out above is that Weyl invariance can be always be restored in Conformal Dilaton Gravity if one is willing to accept counterterms which have logarithmic dependences on the fields. Otherwise, it cannot be restored. Of course, our
analysis rests on the validity of the assumption that the {\em on shell} two-loop UV divergent contribution of Conformal Dilaton Gravity in 4 dimensions can be obtained from the Goroff and Sagnotti counterterm by performing the appropriate Weyl transformation. This hypothesis is suggested
by our one loop results, but, obviously, it demands to be confirmed, or  falsified, by carrying out the corresponding two loop computation.

Now we come to the case of CDG in 6 dimensions. The discussion parallels thoroughly the discussion carried out in the case in 4 dimensions, but now
the substractions are one loop. Indeed, Peter van Nieuwenhuizen \cite{PVN}, in a brilliant paper, computed the on shell one loop pole arising in  General Relativity in 6 dimension as early as in 1976. His result reads
\be\label{ctpvn}
\Gamma_\infty^{(PVN)}[\bar{G}_{\m\n}]=-{1\over n-6}{9\over 1120}{1\over 32\pi^2}~\int d^6 x \sqrt{|\bar{G}|} ~W_6^{(6)}[\bar{G}_{\mu\nu}],
\ee
where
\be
W_6^{(6)}\equiv W^{(6)\,\a_1\a_2\a_3\a_4}~ W^{(6)}_{a_3\a_4\a_5\a_6}~ W^{(6)\,a_5\a_6}\,_{\a_1\a_2}.
\ee
The symbol $W^{(6)}_{\mu_1\mu_2\mu_3\mu_4}$ stands for the Weyl tensor -see (\ref{weylten})-- for the metric $\bar{G}_{\mu\nu}$ for $n=6$.
The contribution in (\ref{ctpvn}) is Weyl invariant, so that the Weyl transformation 
 \be
 \bar{G}_{\m\n}=\frac{1}{M_p^2}\,\frac{1}{\sqrt{10}}\,\bp\,\bar{g}_{\m\n}
 \ee
 leaves its form unchanged. Hence, we shall assume that the pole part, at n=6, of one loop dimensionally regularized {\em on shell} background field effective action, $\Gamma[\bar{g}_{\m\n},\bp;\,n]$, of CDG runs thus
\be
\Gamma_\infty^{n=6}[\bar{g}_{\m\n},\bp]=-{1\over n-6}{9\over 1120}{1\over 32\pi^2}~\int d^6 x \sqrt{|\bar{g}|} ~W_6^{(6)}[\bar{g}_{\mu\nu}].
\ee
Notice that unlike the case in 4 dimensions, which we analysed above, the pole part in local, ie, an integrated polynomial of the fields and their derivatives.

The substraction in the case at hand that is analogous to the substraction in (\ref{nmaction}) reads
\be\label{nmaction6}
\Gamma_{0}^{(nm,\,n=6)}[\bar{g}_{\mu\nu},\bp]=\lim_{n\rightarrow 6}\{\,\Gamma[\bar{g}_{\mu\nu},\bp;\,n]- \frac{1}{n-6}\,{\cal W}_{-1}^{(nm)}[\bar{g}_{\mu\nu},\bp]\},
\ee
where
\be
{\cal W}_{-1}^{(nm)}[\bar{g}_{\mu\nu},\bp]=-{9\over 1120}{1\over 32\pi^2}~\int d^n x \sqrt{|\bar{g}|} ~W_6^{(n)}[\bar{g}_{\mu\nu}].
\ee
$W_6^{(n)}[\bar{g}_{\mu\nu}]$ is defined in (\ref{wsixn}). 

Now, taking into account (\ref{regwardid}) and the following equation 
\be
\Big[2\,\bar{g}^{\mu\nu}\,{\delta\phantom{\bar{g}^{\mu\nu}}\over\delta\bar{g}^{\mu\nu}}\big]{\cal W}_{-1}^{(nm)}[\bar{g}_{\mu\nu},\bp]=
-(n-6)\,{\cal W}_{-1}^{(nm)}[\bar{g}_{\mu\nu},\bp],
\ee
one shows that $\Gamma_{0}^{(nm,\,n=6)}[\bar{g}_{\mu\nu},\bp]$ in (\ref{nmaction6}) satisfies the following broken Ward identity
\be
\Big[2\,\bar{g}^{\mu\nu}\,{\delta\phantom{\bar{g}^{\mu\nu}}\over\delta\bar{g}^{\mu\nu}}+2\bp\,{\delta\phantom{\bp}\over\delta\bp}\,\Big]
\Gamma_0^{(nm,\,n=6)}[\bar{g}_{\mu\nu},\bp]\,=\,{\cal B}[\bar{g}_{\mu\nu},\bp]={\cal W}_{-1}[\bar{g}_{\mu\nu},\bp].
\ee
${\cal W}_{-1}[\bar{g}_{\m\n},\bp]$ is given by
\be
{\cal W}_{-1}[\bar{g}_{\mu\nu},\bp]=-{9\over 1120}{1\over 32\pi^2}~\int d^4 x \sqrt{|\bar{g}|} ~W_6^{(4)}[\bar{g}_{\mu\nu}].
\ee

As in the case in 4 dimensions,  we introduce next a substraction term that is Weyl invariant in $n$ dimensions:
\be\label{infinitywi6}
\Gamma_\infty^{(wi6)}[\bar{g}_{\mu\nu},\bp;\,n]=-\frac{1}{n-6}
{9\over 1120}{1\over32\pi^2}\frac{1}{M_p^{n-6}}\Big(\frac{n-2}{8(n-1)}\Big)^{\frac{n-6}{n-2}}
\int d^n x \sqrt{|\bar{g}|}\,   \big(\bp\big)^{2\frac{n-6}{n-2}}~W_6^{(n)}[\bar{g}_{\m\n}].
\ee
This substraction term is obtained by applying the Weyl transformation in (\ref{nweyltrans}) to the geometrically natural generalization of $\Gamma_\infty^{(PVN)}[\bar{G}_{\m\n}]$ in (\ref{ctpvn}). 

With the help of $\Gamma_\infty^{(wi6)}[\bar{g}_{\mu\nu},\bp;\,n]$, we define a renormalized action in 6 dimensions that is Weyl invariant as
follows
\be
\Gamma_{0}^{(wi6)}[\bar{g}_{\mu\nu},\bp]=\lim_{n\rightarrow 6}\{\,\Gamma[\bar{g}_{\mu\nu},\bp;\,n]-
\Gamma_\infty^{(wi6)}[\bar{g}_{\mu\nu},\bp;\,n]\}.
\ee

Some comments are now in order. First, in the limit $n\rightarrow 6$, the difference between the non-meromorphic substraction term $\Gamma_\infty^{(wi6)}[\bar{g}_{\mu\nu},\bp;\,n]$, in (\ref{infinitywi6}), and the polynomial substraction $1/(n-6){\cal W}_{-1}^{(nm)}[\bar{g}_{\mu\nu},\bp]$, in (\ref{nmaction6}), contains the UV finite non-meromorphic term
\be
{\cal C}=-{9\over 1120}{1\over 32\pi^2}~\int d^4 x \sqrt{|\bar{g}|}~\big(\ln\sqrt{\bp}\big) ~W_6^{(4)}[\bar{g}_{\mu\nu}].
\ee

It can be readily seen that if we substract the previous  UV finite term, ${\cal C}$, to the renormalized action, $\Gamma_{0}^{(nm,\,n=6)}[\bar{g}_{\mu\nu},\bp]$, in (\ref{nmaction6}), one obtains a new renormalized action that is Weyl invariant.
All this is in complete analogy with the case in 4 dimensions, analysed above. There is however a conspicuous difference: The pole part in 6 dimensions is local, so to modify the non-local structure of the Green function may clash with general principles of quantum field theory, such as unitarity. All these issues deserve to be carefully studied on their own.

\section{Conclusions.}
The conformal invariant action analyzed in our paper (CDG) has been argued \cite{Henz} to be related to the ultraviolet fixed point  of the exact renormalization group equations (ERGE). The general action they considered was
\be
S=\int d(vol)\left(V(\phi^2)+F(\phi^2)~R+{1\over 2}~g^{\m\n}\nabla_\m\phi\nabla_\n\phi\right)
\ee

We find this result quite remarkable, although our results indicate that CDG is not stable under renormalization at least perturbatively.
\par
The main conjecture of the present paper is that the conformal Ward identity is violated  in renormalized CDG at two loops even on shell, if counterterms involving logarithms of the scalar fiels are not allowed and that this Ward identity can be restored if those counterterms are accepted as valid to define a quantum theory of gravity. This is true provided the CDG counterterm can be obtained on shell from the corresponding counterterm in GR. We have proved explicitly this at the one loop order through a not altogether completely trivial calculation, and it is natural to assume that it holds true also to two loops, but we have no proof of this. As to whether this violation ought to be called an anomaly, we are aware that this concept is slippery when dealing with a theory which is not renormalizable, so that new counterterms are expected to appear at any new loop order in the computation.
\par
In spite of the fact that the one loop counterterm vanishes on shell, its rather intricate off shell structure should affect computations other than S-matrix ones.
\par
An interesting topic that we did not touch in this paper is the analysis of the theory in the symmetric phase. Background field techniques fail in this case owing to the fact that there is no propagator for the gravitational fluctuations. One could modify the action by introducing an Einstein-Hilbert piece
\be
\m^2 \int d(vol) R
\ee
and then take the limit when $\m^2\rightarrow 0$. It seems that this is equivalent to a constant value for the classical scalar field, namely,
\be
\bp=\m
\ee
Then the counterterm reduces in this case to just three terms independent of $\m$; using our previous notation
\bea
&&a_{2}\left[\bg,\bp\right]=720 \left(-1 + 12 \Ep\right) \left\{-P_7(\xi,\g)\br_{\m\n}\br^{\m\n}+P_{10}(\xi,\g)\br^{2}+ P_{11}(\xi,\g) \br_{\m\n\a\b}\br^{\m\n\a\b}\right\}\nonumber\\
&
\eea
The theory is never conformal when $\m\neq 0$, so that it seems diffcult to reach the conformal point using this procedure.
Another possibility is to introduce a propagator for the gravitons through gauge fixing. This resembles some aspects of the quantization of topological field theories \cite{Giavarini}. We are planning to continue thinking on this fascinating problem and hope to be able to report on it in the future.
\par
At any rate, it would be most interesting to study the behavior of matter added so that the resulting lagrangian is still  conformal.
\par
It is well known that the Goroff-Sagnotti counterterm does not have any supersymmetric extension. Our arguments therefore do not stand for the supersymmetric extension of CDG, which is also a conformal supergravity, which could well be all-order anomaly-free.

\section*{Acknowledgments}
E.A. has enjoyed many discussions and/or correspondence with Luis Alvarez-Gaum\'e, Stanley Deser, Michael Duff, Renata Kallosh, Andrei Linde, Sergei Odintsov  and Roberto Percacci. M. H-V. acknowledges discussions with Itzhak Bars,  Diego Blas, Victor Martin-Lozano, Miguel Montero, Juan Miguel Nieto,  Guillem P\'erez-Nadal, Alberto Salvio, Sergei Sibiryakov,  and Alessandro Strumia. We are also grateful to Alexander Kamenshchik and Ilya Shapiro for informative emails on their work.  M. H-V. wishes to thank the Theory Unit at CERN for kind hospitality during the final stages of this work. We have been partially supported by the European Union FP7  ITN INVISIBLES (Marie Curie Actions, PITN- GA-2011- 289442)and (HPRN-CT-200-00148) as well as by FPA2009-09017 (DGI del MCyT, Spain), FPA2011-24568 (MICINN, Spain) and S2009ESP-1473 (CA Madrid).  The authors acknowledge the support of the Spanish MINECO {\em Centro de Excelencia Severo Ochoa} Programme under grant  SEV-2012-0249. Algebraic computations have been made with the help of the xAct\cite{Martingarcia} package.

\appendix

\section{A quick reminder of the heat kernel approach.}\label{apendixA}

Let us define now the {\em heat kernel} associated to the operator whose determinant we want to compute as the formal expression
\be
K(\tau)\equiv e^{-\tau \Delta}
\ee

Again formally the inverse operator is given through
\be
\Delta^{-1}(x,y)\equiv \int_0^\infty d\t~K(\t;x,y)
\ee
where the heal kernel obeys the EDP heat equation
\be
\left(\frac{\pd}{\pd\tau}+\Delta_x\right)K(x,y;\tau)=0
\ee
with the boundary condition
\be
K(x,y,0)=\delta^{(n)}(x-y)
\ee

Then
\be
\Delta_x\int_0^\infty K(\t;x,y)=-\int_0^\infty d\t~{\pd\over \pd\tau}K(\t,x,y)=\d^{n}(x-y)
\ee

The class of operators that have been studied by mathematicians \cite{Gilkey} are deformations of the laplacian  of the type
\be
\Delta\equiv D^{\m}D_{\m}+Y
\ee
where the gauge covariant derivative is given by
\be
D_{\m}\equiv \pd_{\m}+X_{\m}
\ee

In the particular case $X=Y=0$ the flat space solution is given by
\be
K_0(x,y;\tau)=\frac{1}{(4\pi \tau)^{n/2}}e^{-\frac{2 \sigma\left(x,y\right)}{4\tau}}
\ee
where the geodesic distance in flat space is simply
\be
\sigma\left(x,y\right)\equiv {1\over 2}~(x-y)^2
\ee

It is clear from the above expression that when
\be
\s\rightarrow 0
\ee
the dominant terms in the above expression will be given by
\be
\t\sim 0
\ee

It is customary in the literature to dub $\t$ as {\em proper time}, although it has really dimensions, of length squared. It is then physically reasonable that the UV behavior of the theory is captured by the corresponding behavior of the heat kernel when $\t\sim 0$. This is fortunate, because there is a beautiful geometrical way of systematically studying this behavior. Besides, the computations are well adapted to general riemannian backgrounds. This method is currently the easiest and most powerful way of getting the divergent piece of the effective action in gauge theories with nontrivial backgrounds.
\par
The simplest approach to get small proper time expansion is due to  Schwinger and Dewitt and simply consists in a brute force Taylor expension
\be
K(\tau;x,y)=K_0 (\tau;x,y)\sum_{p=0}~a_p (x,y)~\tau^p
\ee
with the diagonal part of the first coefficient normalized to 1
\be
a_0(x,x)=1
\ee

The integrated coefficients will be denoted by capital letters
\be
A_n\equiv \int \sqrt{|g|}~d^n x~ a_n(x,x)
\ee
in such a way that
\be
A_0=vol
\ee

The determinant of the operator is then given by
\be
\log\det\Delta\equiv -\int\frac{d\tau}{\tau} \text{tr}~K(\tau)\equiv -\lim_{\s\rightarrow 0}\int_0^{\infty}\frac{d\tau}{\tau}\frac{1}{(4\pi \tau)^{n/2}}\sum_{p=0}^\infty\tau^p \text{tr}~ a_p(x,x)~ e^{-\frac{\sigma^2}{4\tau}}
\ee
where we have regularized the determinant by point-splitting the points $x$ and $y$ (although still keeping only the diagonal part on the small time coefficients).
All ultraviolet divergences are given by the behavior in the $\t\sim 0$ endpoint. The Schwinger-de Witt expansion leads to
\be
\text{log~det}~\Delta=-\sum_{p=0}^\infty \frac{ \sigma^{2p-n}}{4^p \pi^{n\over 2}}~\Gamma\left({n\over 2}-p\right)~\text{tr}\, a_p(x,x)
\ee

The term $p=0$ diverges whn $\s\rightarrow 0$ in four dimensions as
\be
{1\over \s^4}
\ee
but this divergence  is common to all operators and can be absorbed into the cosmological constant. The next term corresponds to $p=2$, and is independent on $\s$. When $n=4-\e$  is given by
\be
\text{log~det}\left.\Delta\right|_{n=4}\equiv\frac{1}{16\pi^2 \left(n-4\right)}~a_2(x,x)
\ee

From this term on, the limit $\s\rightarrow 0$ kills everything.
\par
There are of course finite contributions that are not captured by the small proper time expansion; those are much more difficult to compute and the heat kernel method is not particularly helpful in that respect.
\par
A different way of doing things is by considering $\s=0$ from the very beginning, but including a lower end cutoff, ${1\over \Lambda^2}$ in the proper time integral. It should be remarked that this cutoff respects all symmetries of the theory; it is not a momentum cutoff, and it is therefore compatible with diff as well as conformal invriance. Integrals are extended until an infrared cutoff ${1\over \m^2} $ whcich physically represents the range of validity of the short proper time expansion. The result is

\be
\text{log~det}\left.\Delta\right|_{n=4}\equiv\frac{1}{16\pi^2 }~\left(\Lambda^4 Vol+ {1\over 2}a_1(x,x)\Lambda^2+a_2(x,x)~\text{log}~{\Lambda^2\over \m^2}\right)
\ee

The class of operators we are able to consider are some deformations of the Laplace operator, namely,
\begin{align}
D=-\left(G_{AB}g^{\m\n}\partial_{\m}\partial_{\n}+a_{AB}^{\sigma}\partial_{\sigma}+b_{AB}\right)
\end{align}
where $g^{\m\n}$ is the inverse metric tensor on $M$ and $G_{AB}$  is the metric tensor of the ''gauge" vector bundle $V$ over the space-time manifold $M$, and $a^{\sigma}$ and $b$ are  matrix valued functions on $M$ respectively. Then, there is a unique connection on $V$ and a unique endomorphism $E$ of $V$ so that
\begin{align}
D=-\left(G_{AB} g^{\m\n}D_{\m}D_{\n}+E_{AB}\right)
\end{align}
where the covariant derivative $D=\D+\omega$ contains both Riemann and ''gauge" bundle parts. The introduction of the bundle with capital indices will allow us to encode the collection of different fields present in our action in a compact structure.
\par
The divergent part of the one-loop effective action in four dimensions is then
\be
{\cal W}_{(1)}=\frac{1}{n-4}\left.A_{2}\right|_{n=4}
\ee

Furthermore, there is an explicit formula for this coefficient, namely
 \begin{align}\label{heat kernel}
A_{2}=\frac{1}{360(4\pi)^{\frac{n}{2}}}\int d^{n}x\sqrt{|g|}\;Tr\left[ 60 R E + 180 E^{2} +5 R^2 - 2 R_{\m\n}R^{\m\n} +2 R^{\m\n\a\b}R_{\m\n\a\b}+30 F_{\m\n}F^{\m\n}\right]
\end{align}
where $F_{\m\n}$ is the field strenght defined by Ricci's identity as
\be
\left[D_{\m},D_{\n}\right]\Psi^{A}=F_{\m\n B}^{A}\Psi^{B}
\ee
 $\Psi^{B}$ being a vector field living on $M$, a section of the vector bundle. The trace refers both to spacetime indices and bundle capital indices.

\section{Some details on the computation. }\label{apendixB}
Including up to quadratic order in the quantum fluctuations we get a quite involved expression, namely
\begin{align}
S_{2}=&-\int d^{n}x\sqrt{|\bg|}\; (H+F+HF)
\end{align}
where
\begin{align*}
H=&\xi\bf^{2}\left[-\frac{1}{4}h\bD^{2}h+\frac{1}{4}h^{\a\b}\bD^{2}h_{\a\b}-\frac{1}{2}\bD_{\m}h\bD_{\n}h^{\m\n}+\frac{1}{2}\bD_{\m}h^{\m\a}\bD_{\n}h^{\n}_{\a}+ \frac{1}{8}h^{2}\br-\frac{1}{4}h_{\m\n}h^{\m\n}\br-\right.\\
&\left. -\frac{1}{2}h h^{\a\b}\br_{\a\b}+\frac{1}{2}h^{\m\n}h^{\a}_{\n}\br_{\m\a}+\frac{1}{2}\br_{\m\n\a\b}h^{\m\a}h^{\n\b}\right]+\xi(\bD_{\a}\bf^{2})\left(\frac{1}{4}h\bD^{\a}h-\frac{3}{4}h^{\m\n}\bD^{\a}h_{\m\n}+\right.\\
&\left.  +\frac{3}{2}h^{\a\b}\bD_{\m}h_{\b}^{\m}+\frac{1}{2}h^{\m\n}\bD_{\m}h^{\a}_{\n}-h^{\a\b}\bD_{\b}h-\frac{1}{2}h\bD_{\b}h^{\a\b}\right)+\frac{1}{2}h^{\m\a}h_{\a}^{\n}\bD_{\m}\bf\bD_{\n}\bf -\frac{1}{4}h h^{\m\n}\bD_{\m}\bf \bD_{\n}\bf+\\
&-\frac{1}{8}h_{\m\n}h^{\m\n}\bD_{\a}\bf\bD^{\a}\bf+\frac{1}{16}h^{2}\bD_{\m}\bf\bD^{\m}\bf\\
\\
F=&\frac{1}{2}\bD^{\a}\f\bD_{\a}\f + \xi \br\f^{2}\\
\\
HF=&\xi\bf\phi\left(-2h^{\m\n}\br_{\m\n}+h\br+2\bD_{\m}\bD_{\n}h^{\m\n}-2\bD^{2}h\right)-h^{\a\b}\bD_{\a}\f\bD_{\b}\bf +\frac{1}{2} h \bD_{\m}\bf \bD^{\m}\f
\end{align*}

Since gravitational fluctuations are symmetric tensors, $h_{\m\n}=h_{\n\m}$ only the symmetric part of the quadratic term contributes. We found it convenient to define the operators
 \begin{align}
{\cal P}_{\m\n\r\s}^{\a\b}&=\frac{1}{8}\left(\bg_{\m\r}\delta_{\n}^{\a}\delta_{\s}^{\b}+\bg_{\m\s}\delta^{\a}_{\n}\delta^{\b}_{\r}+\bg_{\n\r}\delta^{\a}_{\m}\delta^{\b}_{\s}+\bg_{\n\s}\delta^{\a}_{\m}\delta^{\b}_{\r}\right)+\frac{1}{8}(\alpha \leftrightarrow \beta)\\
\nonumber {\cal K}_{\m\n\r\s}^{\a\b}&=\frac{1}{4}\left(\bg_{\m\n}\delta^{\a}_{\r}\delta^{\b}_{\s}+\bg_{\r\s}\delta^{\a}_{\m}\delta^{\b}_{\n}\right)+\frac{1}{4}(\alpha\leftrightarrow\beta)
\end{align}
It is then plain that
\bea
&&h^{\m\n}h_{\m\n}=h^{\m\n}h^{\r\s}{\cal P}_{\m\n\r\s}^{\a\b}\bg_{\a\b}\nonumber\\
&&h^{2}={\cal K}_{\m\n\r\s}^{\a\b}\bg_{\a\b}h^{\m\n}h^{\r\s}
\eea

After introducing the preceding notation, the quadratic operators read
\begin{align*}
\hat{H}_{\m\n\r\s}=&\xi\bf^{2}\left[\frac{1}{4}\left({\cal P}_{\m\n\r\s}^{\a\b}-{\cal K}_{\m\n\r\s}^{\a\b}\right)\bg_{\a\b}\bD^{2}+\frac{1}{2}\left({\cal P}_{\m\n\r\s}^{\a\b}-{\cal K}_{\m\n\r\s}^{\a\b}\right)\br_{\a\b}+\frac{1}{2}\br_{(\m\r\n\s)}\right.\nonumber \\
\nonumber&\left. +\left(\frac{1}{8}{\cal K}_{\m\n\r\s}^{\a\b}-\frac{1}{4}{\cal P}_{\m\n\r\s}^{\a\b}\right)\bg_{\a\b}\br\right]+\left(\frac{1}{2}{\cal P}^{\a\b}_{\m\n\r\s}-\frac{1}{4}{\cal K}^{\a\b}_{\m\n\r\s}\right)\left(\bD_{\a}\bf \bD_{\b}\bf - \frac{1}{4}\bg_{\a\b}(\bD\bf)^{2}\right)+\\
&+\frac{\xi}{4}(\bD_{\a}\bf^{2})\left(\left({\cal K}_{\m\n\r\s}^{\gamma\omega}-3{\cal P}_{\m\n\r\s}^{\gamma\omega}\right)\bg_{\gamma\omega}\bg^{\a\b}+X_{\m\n\r\s}^{\alpha\b}\right)\bD_{\b}
\end{align*}
with the tensor $X_{\m\n\r\s}^{\alpha\b}$ defined as
\begin{align*}
X^{\a\b}_{\m\n\r\s}&=\frac{3}{2}\left(\bg^{\a}_{\m}\bg_{\r}^{\b}\bg_{\n\s}+\bg^{\a}_{\n}\bg^{\b}_{\r}\bg_{\m\s}+\bg^{\a}_{\m}\bg^{\b}_{\s}\bg_{\n\r}+\bg^{\a}_{\n}\bg^{\b}_{\s}\bg_{\m\r}\right)-\left(\bg_{\r}^{\a}\bg^{\b}_{\s}\bg_{\m\n}+\bg^{\a}_{\s}\bg^{\b}_{\r}\bg_{\m\n}\right)+\\
&+\frac{1}{2}\left(\bg^{\a}_{\r}\bg^{\b}_{\m}\bg_{\n\s}+\bg^{\a}_{\r}\bg^{\b}_{\n}\bg_{\m\s}+\bg^{\a}_{\s}\bg^{\b}_{\m}\bg_{\n\r}+\bg^{\a}_{\s}\bg^{\b}_{\n}\bg_{\m\r}\right)-2\left(\bg^{\a}_{\m}\bg^{\b}_{\n}\bg_{\r\s}+\bg^{\a}_{\n}\bg^{\b}_{\m}\bg_{\r\s}\right)
\end{align*}
and
\begin{align*}
(\widehat{HF})_{\m\n}&=\xi\bf\left[\br \bg_{\m\n}-2\br_{\m\n}+2\bD_{\m}\bD_{\n}-2\bg_{\m\n}\bD^{2}\right]+\frac{1}{2}\left(\bD_{\m}\bf\bD_{\n}+\bD_{\n}\bf\bD_{\m}\right)-\\
\nonumber&-\frac{1}{2}\bg_{\m\n}\bD_{\a}\bf \bD^{\a}+\bD_{\m}\bD_{\n}\bf-\frac{1}{2}\bg_{\m\n}\bD^{2}\bf\\\
\nonumber \\
\hat{F}&=-\frac{1}{2}\bD^{2}+\xi \br
\end{align*}

The operators after the rescaling to the $k_{\m\n}$ variables read
\begin{align*}
\hat{H}_{\m\n\r\s}=&\xi \left[\frac{1}{4}\left({\cal P}_{\m\n\r\s}^{\a\b}-{\cal K}_{\m\n\r\s}^{\a\b}\right)\bg_{\a\b}\bD^{2}+\frac{1}{2}\left({\cal P}_{\m\n\r\s}^{\a\b}-{\cal K}_{\m\n\r\s}^{\a\b}\right)\br_{\a\b}+\frac{1}{2}\br_{(\m\r\n\s)}\right. \\
\nonumber&\left. +\left(\frac{1}{8}{\cal K}_{\m\n\r\s}^{\a\b}-\frac{1}{4}{\cal P}_{\m\n\r\s}^{\a\b}\right)\bg_{\a\b}\br\right]+\left(\frac{1}{2}{\cal P}^{\a\b}_{\m\n\r\s}-\frac{1}{4}{\cal K}^{\a\b}_{\m\n\r\s}\right)\left(\frac{\bD_{\a}\bf \bD_{\b}\bf}{\bf^{2}} - \frac{1}{4}\bg_{\a\b}\frac{(\bD\bf)^{2}}{\bf^{2}}\right)+\\
\nonumber &+2\xi \frac{\bD_{\a}\bf}{\bf}\left(\left(\frac{1}{2}{\cal K}_{\m\n\r\s}^{\gamma\omega}-{\cal P}_{\m\n\r\s}^{\gamma\omega}\right)\bg_{\gamma\omega}\bg^{\a\b}+\frac{1}{2}{\cal K}_{\m\n\r\s}+\frac{1}{4}Y_{\m\n\r\s}^{\alpha\b}\right)\bD_{\b}+\\
\nonumber&+\frac{\xi}{2}\left[\frac{1}{2}\left({\cal P}_{\m\n\r\s}^{\a\b}-{\cal K}_{\m\n\r\s}^{\a\b}\right)\bg_{\a\b}\left(2\frac{(\bD\bf)^{2}}{\bf^{2}}  - \frac{\bD^{2}\bf}{\bf}\right)+\right.\\
\nonumber&\left. +\left({\cal P}_{\m\n\r\s}^{\a\b}-{\cal K}_{\m\n\r\s}^{\a\b}-\left({\cal K}_{\m\n\r\s}^{\gamma\delta}-3{\cal P}_{\m\n\r\s}^{\gamma\delta}\right)\bg_{\gamma\delta}\bg^{\a\b} - X^{\a\b}_{\m\n\r\s}\right)\frac{\bD_{\a}\bf\bD_{\b}\bf}{\bf^{2}}\right]
\end{align*}
\begin{align*}
(\widehat{HF})_{\m\n}&=\xi\left[\br \bg_{\m\n}-2\br_{\m\n}+2\bD_{\m}\bD_{\n}-2\bg_{\m\n}\bD^{2}\right]+\\
\nonumber &+\frac{1}{2}\left(\frac{\bD_{\m}\bf}{\bf}\delta^{\b}_{\n}+\frac{\bD_{\n}\bf}{\bf}\delta^{\b}_{\m}-\bg_{\m\n}\frac{\bD^{\b}\bf}{\bf}\right)\bD_{\b}-\\
\nonumber &-2\xi \left(\frac{\bD_{\m}\bf}{\bf}\delta^{\b}_{\n}+\frac{\bD_{\n}\bf}{\bf}\delta^{\b}_{\m}-2\bg_{\m\n}\frac{\bD^{\b}\bf}{\bf}\right)\bD_{\b}+\\
\nonumber &+2\xi \left(2\frac{\bD_{\m}\bf\bD_{\n}\bf}{\bf^{2}}-\frac{\bD_{\m}\bD_{\n}\bf}{\bf}-2\frac{(\bD\bf)^{2}}{\bf^{2}}\bg_{\m\n}+\frac{\bD^{2}\bf}{\bf}\bg_{\m\n}\right)+\\
\nonumber &+\frac{1}{2}\left(2\frac{\bD_{\m}\bD_{\n}\bf}{\bf}-2\frac{\bD_{\m}\bf\bD_{\n}\bf}{\bf^{2}}+\frac{(\bD\bf)^{2}}{\bf^{2}}-\bg_{\m\n}\frac{\D^{2}\bf}{\bf}\right)\\
\nonumber \\
\hat{F}&=-\frac{1}{2}\bD^{2}+\frac{(n-2)}{8(n-1)}\br
\end{align*}
and finally
\begin{align*}
Y^{\a\b}_{\m\n\r\s}&=\left(\bg^{\a}_{\m}\bg_{\r}^{\b}\bg_{\n\s}+\bg^{\a}_{\n}\bg^{\b}_{\r}\bg_{\m\s}+\bg^{\a}_{\m}\bg^{\b}_{\s}\bg_{\n\r}+\bg^{\a}_{\n}\bg^{\b}_{\s}\bg_{\m\r}\right)-\left(\bg_{\r}^{\a}\bg^{\b}_{\s}\bg_{\m\n}+\bg^{\a}_{\s}\bg^{\b}_{\r}\bg_{\m\n}\right)+\\
\nonumber &+\frac{1}{2}\left(\bg^{\a}_{\r}\bg^{\b}_{\m}\bg_{\n\s}+\bg^{\a}_{\r}\bg^{\b}_{\n}\bg_{\m\s}+\bg^{\a}_{\s}\bg^{\b}_{\m}\bg_{\n\r}+\bg^{\a}_{\s}\bg^{\b}_{\n}\bg_{\m\r}\right)-2\left(\bg^{\a}_{\m}\bg^{\b}_{\n}\bg_{\r\s}+\bg^{\a}_{\n}\bg^{\b}_{\m}\bg_{\r\s}\right)
\end{align*}

\par

The operators that appear in the gauge fixed action read
\begin{align}
S^{\text{full}}_2=-\int d^{n}x\sqrt{|\bg|}\left[k^{\m\n}\hat{H}_{\m\n\r\s}k^{\r\s}+\phi (\widehat{HF})_{\m\n} k^{\m\n}+\phi\hat{F}\phi\right]
\end{align}
with
\begin{align*}
\hat{H}_{\m\n\r\s}=&\xi\left[\frac{1}{4}\left({\cal P}_{\m\n\r\s}^{\a\b}-\frac{1}{2}{\cal K}_{\m\n\r\s}^{\a\b}\right)\bg_{\a\b}\bD^{2}+\frac{1}{2}\left({\cal P}_{\m\n\r\s}^{\a\b}-{\cal K}_{\m\n\r\s}^{\a\b}\right)\br_{\a\b}+\frac{1}{2}\br_{(\m\r\n\s)}\right. \\
\nonumber&\left. +\left(\frac{1}{8}{\cal K}_{\m\n\r\s}^{\a\b}-\frac{1}{4}{\cal P}_{\m\n\r\s}^{\a\b}\right)\bg_{\a\b}\br\right]+\left(\frac{1}{2}{\cal P}^{\a\b}_{\m\n\r\s}-\frac{1}{4}{\cal K}^{\a\b}_{\m\n\r\s}\right)\left(\frac{\bD_{\a}\bf \bD_{\b}\bf}{\bf^{2}} - \frac{1}{4}\bg_{\a\b}\frac{(\bD\bf)^{2}}{\bf^{2}}\right)+\\
\nonumber &+2\xi\frac{\bD_{\a}\bf}{\bf}\left(\left(\frac{1}{2}{\cal K}_{\m\n\r\s}^{\gamma\omega}-{\cal P}_{\m\n\r\s}^{\gamma\omega}\right)\bg_{\gamma\omega}\bg^{\a\b}+\frac{1}{2}{\cal K}_{\m\n\r\s}+\frac{1}{4}Y_{\m\n\r\s}^{\alpha\b}\right)\bD_{\b}+\\
\nonumber&+\frac{\gamma^{2}\xi}{4}E^{\a\b}_{\m\n\r\s}\frac{\bD_{\a}\bf\bD_{\b}\bf}{\bf^{2}}+\frac{\xi}{2}\left[\frac{1}{2}\left({\cal P}_{\m\n\r\s}^{\a\b}-{\cal K}_{\m\n\r\s}^{\a\b}\right)\bg_{\a\b}\left(2\frac{(\bD\bf)^{2}}{\bf^{2}}  - \frac{\bD^{2}\bf}{\bf}\right)+\right.\\
\nonumber&\left. +\left({\cal P}_{\m\n\r\s}^{\a\b}-{\cal K}_{\m\n\r\s}^{\a\b}-\left({\cal K}_{\m\n\r\s}^{\gamma\delta}-3{\cal P}_{\m\n\r\s}^{\gamma\delta}\right)\bg_{\gamma\delta}\bg^{\a\b} - X^{\a\b}_{\m\n\r\s}\right)\frac{\bD_{\a}\bf\bD_{\b}\bf}{\bf^{2}}\right]-\frac{\gamma \xi}{2}E^{\a\b}_{\m\n\r\s}\frac{\bD_{\a}\bf}{\bf}\bD_{\b}
\end{align*}
\begin{align*}
(\widehat{HF})_{\m\n}&=\xi\left[\br \bg_{\m\n}-2\br_{\m\n}-\bg_{\m\n}\bD^{2}\right]+\\
\nonumber &+\left(\gamma\xi+\frac{1}{2}\right)\left(\frac{\bD_{\m}\bf}{\bf}\delta^{\b}_{\n}+\frac{\bD_{\n}\bf}{\bf}\delta^{\b}_{\m}-\bg_{\m\n}\frac{\bD^{\b}\bf}{\bf}\right)\bD_{\b}-\\
\nonumber &-2\xi \left(\frac{\bD_{\m}\bf}{\bf}\delta^{\b}_{\n}+\frac{\bD_{\n}\bf}{\bf}\delta^{\b}_{\m}-2\bg_{\m\n}\frac{\bD^{\b}\bf}{\bf}\right)\bD_{\b}+\\
\nonumber &+2\xi \left(2\frac{\bD_{\m}\bf\bD_{\n}\bf}{\bf^{2}}-\frac{\bD_{\m}\bD_{\n}\bf}{\bf}-2\frac{(\bD\bf)^{2}}{\bf^{2}}\bg_{\m\n}+\frac{\bD^{2}\bf}{\bf}\bg_{\m\n}\right)+\\
\nonumber &+\frac{1}{2}\left(2\frac{\bD_{\m}\bD_{\n}\bf}{\bf}-2\frac{\bD_{\m}\bf\bD_{\n}\bf}{\bf^{2}}+\frac{(\bD\bf)^{2}}{\bf^{2}}-\bg_{\m\n}\frac{\bD^{2}\bf}{\bf}\right)\\
\nonumber \\
\hat{F}&=\left(2\xi-\frac{1}{2}\right)\bD^{2}+\xi \br
\end{align*}
\par
where we have introduced
\begin{align*}
E_{\m\n\r\s}^{\a\b}=&\frac{1}{2}\left(g_{\m\n}\delta^{\a}_{\r}\delta^{\b}_{\s}+g_{\m\n}\delta^{\a}_{\s}\delta^{\b}_{\r}+g_{\r\s}\delta^{\a}_{\m}\delta^{\b}_{n}+g_{\r\s}\delta^{\a}_{\n}\delta^{\b}_{m}-g_{\m\n}g_{\r\s}g^{\a\b}\right)-\\
&-\frac{1}{2}\left(g_{\m\r}\delta^{\a}_{\n}\delta^{\b}_{\s}+g_{\m\s}\delta^{\a}_{\n}\delta^{\b}_{\r}+g_{\n\r}\delta^{\a}_{\m}\delta^{\b}_{\s}+g_{\n\s}\delta^{\a}_{\m}\delta^{\b}_{\r}\right)
\end{align*}

The matrix $N^{\b}_{AB}$  reads
\begin{align}
N^{\b}_{AB}=\begin{pmatrix}
N^{\b}_{kk}& N^{\b}_{k\phi}\\
N^{\b}_{\f k}& N^{\b}_{\f\f}
\end{pmatrix}
\end{align}
where
\begin{align*}
&N^{\b}_{kk}= \frac{\xi}{4}\left(Y^{\a\b}_{\m\n\r\s}-Y^{\a\b}_{\r\s\m\n}-\gamma E^{\a\b}_{\m\n\r\s}+\gamma E^{\a\b}_{\r\s\m\n}\right)\frac{\bD_{\a}\bf}{\bf}\\
&N^{\b}_{\f\f}=0\\
&N^{\b}_{k\f}=-N^{\b}_{\f k}=\-\frac{1}{2}\left(\frac{1}{2}-2\xi\right)\left(\frac{\bD_{\m}\bf}{\bf}\delta^{\b}_{\n}+\frac{\bD_{\n}\bf}{\bf}\delta^{\b}_{\m}\right)-\frac{1}{2}\left(4\xi -\frac{1}{2}\right)\bg_{\m\n}\frac{\bD^{\b}\bf}{\bf}-\\
\nonumber &-\frac{\xi\gamma}{2}\left(\frac{\bD_{\m}\bf}{\bf}\delta^{\b}_{\n}+\frac{\bD_{\n}\bf}{\bf}\delta^{\b}_{\m}-\bg_{\m\n}\frac{\bD^{\b}\bf}{\bf}\right)
\end{align*}
where we had to integrate by parts half of the symmetric part in order to cancel it. This will leave some "residues" that  will be introduced into the potential matrix $M_{AB}$, which reads
\begin{align}
M_{AB}=\begin{pmatrix}
M_{kk}& M_{k\phi}\\
M_{\f k}& M_{\f\f}
\end{pmatrix}
\end{align}
where the different elements are
\small \begin{align*}
\nonumber &M_{kk}=\frac{(n-2)}{8(n-1)}\left[\frac{1}{2}\left({\cal P}_{\m\n\r\s}^{\a\b}-{\cal K}_{\m\n\r\s}^{\a\b}\right)\br_{\a\b}+\frac{1}{4}{\cal G}_{\m\n\r\s}^{\a\b}\bg_{\a\b}\br+\frac{1}{2}\br_{(\m\r\n\s)}\right]-\frac{1}{2}{\cal G}^{\a\b}_{\m\n\r\s}\left(\frac{\bD_{\a}\bf \bD_{\b}\bf}{\bf^{2}} - \frac{1}{4}\bg_{\a\b}\frac{(\bD\bf)^{2}}{\bf^{2}}\right)+\\
\nonumber &+\frac{(n-2)}{16(n-1)}\Bigg\{\frac{1}{2}\left({\cal P}_{\m\n\r\s}^{\a\b}-{\cal K}_{\m\n\r\s}^{\a\b}\right)\bg_{\a\b}\left(2\frac{(\bD\bf)^{2}}{\bf^{2}}-\frac{\bD^{2}\bf}{\bf}\right)+\frac{\bD_{\a}\bf\bD_{\b}\bf}{\bf^{2}}\left[\left({\cal P}_{\m\n\r\s}^{\a\b}-{\cal K}_{\m\n\r\s}^{\a\b}\right)+\right.\\
\nonumber &\left. +\left(3{\cal P}_{\m\n\r\s}^{\g\omega}-{\cal K}_{\m\n\r\s}^{\g\omega}\right)\bg_{\g\omega}\bg^{\a\b}-X^{\a\b}_{\m\n\r\s}\right]\Bigg\}-\frac{(n-2)}{8(n-1)}\bD_{\b}\left(\frac{\bD_{\a}\bf}{\bf}\right)\left[{\cal G}_{\m\n\r\s}^{\g\omega}\bg_{\g\omega}\bg^{\a\b}+\frac{1}{2}{\cal K}_{\m\n\r\s}^{\a\b}+\frac{1}{8}\left(Y^{\a\b}_{\m\n\r\s}+Y^{\a\b}_{\r\s\m\n}\right)\right]+\\
\nonumber &+ \frac{\xi}{8}\left(E^{\a\b}_{\m\n\r\s}+E^{\a\b}_{\r\s\m\n}\right)\left(\gamma\bD_{\b}\left(\frac{\bD_{\a}\bf}{\bf}\right)+\gamma^{2}\frac{\bD_{\a}\bf\bD_{\b}\bf}{\bf^{2}} \right)
\end{align*}
\begin{align*}
\nonumber&M_{k\f}=M_{\f k}=\frac{\xi}{2}\left[\br \bg_{\m	\n}-2\br_{\m\n}\right]+\xi\left(2\frac{\bD_{\m}\bf\bD_{\n}\bf}{\bf^{2}}-\frac{\bD_{\m}\bD_{\n}\bf}{\bf}-2\bg_{\m\n}\frac{\bD_{\b}\bf\bD^{\b}\bf}{\bf^{2}}+\bg_{\m\n}\frac{\bD_{\b}\bD^{\b}\bf}{\bf}\right)+\\
\nonumber &+\frac{1}{4}\left(2\frac{\bD_{\m}\bD_{\n}\bf}{\bf}-2\frac{\bD_{\m}\bf\bD_{\n}\bf}{\bf^{2}}+\bg_{\m\n}\frac{\bD_{\b}\bf\bD^{\b}\bf}{\bf^{2}}-\bg_{\m\n}\frac{\bD_{\b}\bD^{\b}\bf}{\bf}\right)-\frac{1}{2}\left(\frac{1}{2}-2\xi\right)\bD_{\m}\left(\frac{\bD_{\m}\bf}{\bf}\right)-\\
\nonumber&- \frac{1}{4}\left(4\xi-\frac{1}{2}\right)\bg_{\m\n}\bD_{\b}\left(\frac{\bD^{\b}\bf}{\bf}\right)-\frac{\gamma\xi}{2}\left[\frac{1}{2}\bg_{\m\n}\bD_{\b}\left(\frac{\bD^{\b}\bf}{\bf}\right)-\bD_{\m}\left(\frac{\bD_{\n}\bf}{\bf}\right)\right]\\
\nonumber \\
\nonumber &M_{\f\f}=\xi \br
\end{align*}
\normalsize

\end{document}